\documentclass[11pt]{article}
\usepackage[T1]{fontenc}
\usepackage[utf8]{inputenc}
\usepackage{graphicx}
\usepackage{amsmath}
\usepackage{amssymb}
\usepackage{tabularx}
\newcolumntype{Y}{>{\raggedright\arraybackslash}X}
\usepackage[round]{natbib}
\usepackage{float}
\usepackage[letterpaper,margin=1in]{geometry}
\usepackage[font=small,labelfont=bf]{caption}
\usepackage{fancyhdr}
\usepackage{xcolor}
\usepackage[
    colorlinks=true,
    linkcolor=blue!60!black,
    citecolor=blue!60!black,
    urlcolor=blue!60!black,
    pdfauthor={Abdullah A. Fahad et al.},
    pdftitle={A Physics-ML Multi-Fidelity Strategy for Earth System Model Parameter Optimization: A QG Proof-of-Concept}
]{hyperref}

\title{\vspace{-1.25em}\bfseries A Physics--ML Multi-Fidelity Strategy for Earth System Model Parameter Optimization: A QG Proof-of-Concept}

\author{
Abdullah A. Fahad$^{1,*}$, Manmeet Singh$^{2}$, Donifan Barahona$^{1}$,\\
Anton Darmenov$^{1}$, Andrea Molod$^{1}$\\[0.5em]
\footnotesize $^{1}$Global Modeling and Assimilation Office, NASA Goddard Space Flight Center,\\
\footnotesize Greenbelt, MD 20771, USA\\[0.3em]
\footnotesize $^{2}$Department of Earth, Environmental and Atmospheric Sciences, Western Kentucky University,\\
\footnotesize Bowling Green, KY 42101, USA\\[0.5em]
\footnotesize $^{*}$Corresponding author: \href{mailto:a.fahad@nasa.gov}{a.fahad@nasa.gov}
}
\date{\small Preprint}

\begin{document}

\maketitle
\thispagestyle{fancy}

\begin{abstract}
Earth System Models (ESMs) are primary tools that we use to understand weather-to-climate dynamics and produce future projections. However, model accuracy can significantly depend on subgrid-scale parameter sensitivity. Optimizing these parameters is computationally expensive, and resolving the correct nonlinear dynamics in high-dimensional parameter space is not straightforward. Linear optimization methods like Green's Function Optimization (GFO) can be very sample-efficient for optimizing parameters and provide interpretable local sensitivity information; however, they lack the ability to capture nonlinear dynamics. In contrast, non-linear machine learning methods (e.g., Gaussian Processes, Neural Networks) are designed to learn nonlinear dynamics; however, they need a large number of samples to train the model. In this work, we propose a hybrid Physics--ML approach with a multi-fidelity optimization technique to optimize the parameters with high sample efficiency and near-optimum solutions. Using the late-time statistical state of a Quasi-Geostrophic (QG) turbulence model as a proof-of-concept proxy for Earth System Models with tunable subgrid-scale parameters, we propose the workflow in two stages: (1) linear screening with GFO to find a highly sensitive active parameter subset and initial optimized values, and (2) using the selected parameters with a multi-fidelity training approach to fit the ML models and find near-optimal parameter optimization values. In the multi-fidelity stage, the reduced number of parameters from GFO is handed to a non-linear surrogate that first explores the loss landscape of optimized parameters with cheap short simulations, and then deploys expensive full-length simulations only for promising candidates. Comparing seven optimization strategies with a spread of ensemble members, the hybrid methods achieved the highest mean improvements, outperforming the other tested end-to-end pipelines while reaching practical convergence in the least simulation-days. These results demonstrate an end-to-end performance advantage for the tested hybrid pipeline; the present experiments do not separately attribute that advantage to screening, dimensionality reduction, initialization, or fidelity scheduling. This proof-of-concept QG model tuning establishes the viability of the hybrid framework for future application to operational ESM parameter optimization without the need to run a large number of expensive simulations.
\end{abstract}

\section{Introduction}

Earth System Models (ESMs) are critical components of modern weather and climate science. They help us understand how the dynamics of the Earth system work, from short-term weather forecasts to long-term climate predictions. The science community also uses ESMs to understand climate change physics that significantly influence our long-term environmental policy. However, running high-resolution simulation forecasts that represent realistic subgrid-scale processes, such as cloud microphysics, turbulence, and convection, is computationally expensive. As a result, ESMs rely on parameterized subgrid-scale processes. The skill of ESMs fundamentally depends on the fidelity of the parameterizations used in the model. These are often uncertain and require optimization to reproduce realistic climate statistics \citep{Held2005, Schmidt2017, Strobach2022}.

Optimized and accurate parameterization is highly significant. However, the ESM parameter optimization process has historically been described as more of an art than a science, mostly because there is no agreed-upon methodology among top institutions for how to adjust these parameters \citep{Mauritsen2012, Hourdin2017, Schmidt2017, Strobach2022}. For example, NCAR and GFDL rely heavily on expert judgment to balance competing metrics in their ESMs and determine optimized parameter values \citep{Schmidt2017}. In contrast, the Max Planck Institute for Meteorology (MPI-M) emphasizes ``equally valid'' optimizations and has found that they can yield different climate sensitivities \citep{Mauritsen2012}. This manual, heuristic approach is labor-intensive and typically proceeds through trial and error.

Automated optimization techniques use systematic algorithmic approaches and move away from heuristic trial-and-error tuning. For example, linearized Green's-function calibration methods can be computationally efficient and provide physically interpretable local sensitivity information \citep{menemenlis2005using, Strobach2022, Carroll2020}. Their local linearization, however, does not represent strongly nonlinear parameter interactions. In contrast, nonlinear machine-learning (ML) methods developed in recent years (e.g., Gaussian Processes (GP), Neural Networks) offer greater flexibility to tune model parameters \citep{Schneider2017}. However, they suffer from the ``curse of dimensionality,'' which can make them prohibitively expensive for high-dimensional models because they are generally not sample efficient.

The use of machine-learning surrogate models is growing in emulator-based approaches to climate-model parameter optimization. For example, \citet{Bonnet2025} used Gaussian-process regression emulators and history matching to constrain ICON-A atmospheric parameters affecting radiation, clouds, and winds from a perturbed-parameter ensemble (PPE). Similarly, \citet{Elsaesser2025} used a calibrated physics ensemble method for the NASA GISS ModelE by training a neural network surrogate on a PPE and embedding it within a Markov chain Monte Carlo framework to infer parameters such as convective entrainment and ice fall speed.

ML models are also applied to tune land-surface and precipitation process by optimizing model parameters. For example, \citet{Dagon2020} used feed forward NNs on a PPE samples of the Community Land Model (CLM5) to produce realistic global carbon and water fluxes by optimized biophysical parameters. More recently, \citet{Wu2025} utilized a multilevel ML surrogate model for tuning precipitation parameters in CAM5. This approach combined Gradient Boosted Regression Trees with local trust-region refinement and achieved an approximate 19\% reduction in regional precipitation errors. These methods explicitly leverage large pre-computed ensembles to train the non-linear surrogates and are not sample efficient by design.

In this study, we propose a novel hybrid multi-fidelity optimization approach that combines the efficiency of linear methods with the precision of non-linear optimization. Our framework proceeds in two stages to maximize sample efficiency. First, we use a linear screening phase with full-length GFO simulations to perform a coarse optimization and identify the subset of ``active'' parameters. Second, we utilize a non-linear refinement phase, where a multi-fidelity surrogate model leverages short simulations to learn the topological structure of the loss landscape. By using an acquisition function to identify the optimization valley, this stage efficiently converges toward a near-optimal solution with minimal high-fidelity evaluations.

\section{Methodology}

\subsection{Quasi-Geostrophic (QG) Turbulence Model Optimization}

To validate our framework, we use the two-level beta-plane turbulence implementation supplied with this study, motivated by classical rotating and beta-plane quasi-geostrophic turbulence literature \citep{DeVerdiere1980, Panayotova2007, Gryanik2004}. We describe the implemented equations explicitly because this simplified proof-of-concept is not a direct implementation of any one of those cited models and is not a complete stratified two-layer QG solver. In the released code, each level is inverted independently according to
\begin{equation}
    \widehat{\psi_i}(\boldsymbol{k})=-\frac{\widehat{q_i}(\boldsymbol{k})}{{|\boldsymbol{k}|^2}},\qquad
    \widehat{\psi_i}(\boldsymbol{0})=0,
\end{equation}
so that $q_i=\nabla^2\psi_i$ for nonzero Fourier modes. The prognostic tendencies before spectral filtering are
\begin{align}
    \frac{\partial q_1}{\partial t} &= -J(\psi_1,q_1)-\beta\frac{\partial\psi_1}{\partial x}
    +10^{-8}\epsilon_{back}\,\xi_1-\epsilon_{ens}q_1,\\
    \frac{\partial q_2}{\partial t} &= -J(\psi_2,q_2)-\beta\frac{\partial\psi_2}{\partial x}
    +\alpha_r r_{ek}\nabla^2\psi_2+10^{-8}\epsilon_{back}\,\xi_2-\epsilon_{ens}q_2,
\end{align}
where $J(a,b)=a_xb_y-a_yb_x$ and the grid-point fields $\xi_i$ are independent standard-normal draws at each time step when $\epsilon_{back}\ne0$. The domain is doubly periodic. The code computes $k_d=[f_0^2/(g'H_1H_2/(H_1+H_2))]^{1/2}$ from the configuration metadata, but $k_d$ does not enter the implemented PV inversion or tendencies. Consequently, the two levels in this implementation are not coupled dynamically; their depths enter the diagnostics and barotropic loss projection. Appendix~\ref{app:qg_impl} provides the complete numerical configuration and filtering procedure.

\begin{figure}[ht]
    \centering
    \includegraphics[width=0.9\textwidth]{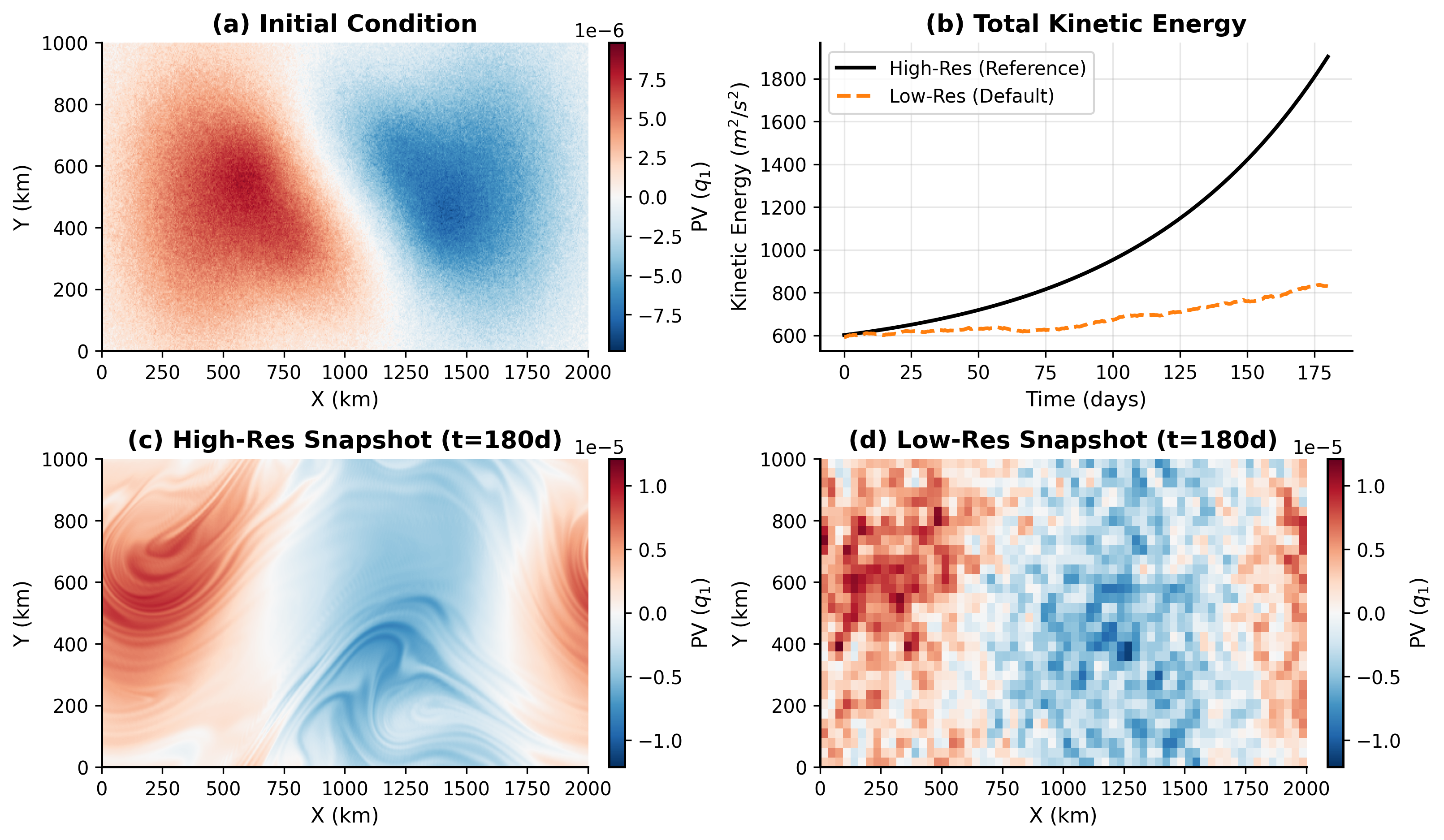}
    \caption{High-Resolution Dynamics. (a) Initial PV Field. (b) Time evolution of Total Kinetic Energy comparing the High-Resolution Reference (Black) against the drifting Low-Resolution Default (Orange). (c) Mature High-Resolution PV Snapshot. (d) Low-Resolution Default PV Snapshot, showing coherent structures but incorrect energy levels and dissipation.}
    \label{fig:qg_dynamics}
\end{figure}

To construct the optimization problem, we define a high-resolution reference configuration and a coarse proxy configuration. The high-resolution case\footnote{High-resolution simulation video: \url{https://www.youtube.com/shorts/pmDJhbrb-0E}} uses a $2000 \times 1000$ km domain resolved on a $512 \times 256$ grid ($\Delta x \approx 3.9$ km). The low-resolution proxy uses the same physical domain on a $64 \times 32$ grid ($\Delta x \approx 31.25$ km), an eightfold coarsening in each horizontal direction. This resolution contrast removes much of the fine-scale vorticity structure and creates a controlled SGS-tuning problem. The configured diagnostic scale $k_d^{-1}\simeq34.9$ km is well resolved by the HR grid and comparable to the LR grid spacing, but it is not an active dynamical scale in the archived uncoupled inversion described above. The HR and LR fields are initialized independently at their native grids using matched ensemble-generation rules rather than by deterministically coarse-graining one common field. The resulting objective therefore measures the aggregate statistical discrepancy between the reference and proxy configurations, including both resolution-dependent evolution and realization-dependent initial adjustment; it should not be interpreted as a pure discretization-error measure.

The low-resolution configuration exposes six candidate controls. Viscosity Scale ($\alpha_{\nu}$) multiplies the coefficient of the eighth-order spectral filter, and Linear Drag Scale ($\alpha_r$) multiplies lower-level Ekman drag. Eddy Diffusivity ($K_{eddy}$) enters a separate biharmonic spectral filter, Energy Correction ($\epsilon_{back}$) multiplies the grid-point Gaussian term above, and Enstrophy Correction ($\epsilon_{ens}$) is linear PV damping. Although the source defines a routine that diagnoses a Smagorinsky viscosity from $C_s$, that routine is not called by the time-stepping code in the archived implementation. Thus $C_s$ has no effect on these simulations and should be regarded as an inactive code parameter rather than an implemented SGS tendency. We retain it in the screening exercise as a null-control check.

\begin{table}[H]
    \centering
    \caption{Subgrid-Scale Parameter Optimization Ranges}
    \label{tab:sgs_params}
    \begin{tabular}{lccc}
        \hline
        \textbf{Parameter} & \textbf{Symbol} & \textbf{Range} & \textbf{Unit} \\
        \hline
        Viscosity Scale & $\alpha_{\nu}$ & $0.5 - 5.0$ & Non-dim \\
        Linear Drag Scale & $\alpha_{r}$ & $0.5 - 3.0$ & Non-dim \\
        Eddy Diffusivity & $K_{eddy}$ & $10^3 - 10^5$ & $m^4/s$ \\
        Smagorinsky Coeff. & $C_s$ & $0.0 - 0.3$ & Non-dim \\
        Stochastic Correction Scale & $\epsilon_{back}$ & $-0.01 - 0.01$ & Code scale \\
        Enstrophy Correction & $\epsilon_{ens}$ & $0.0 - 10^{-6}$ & $s^{-1}$ \\
        \hline
    \end{tabular}
\end{table}

The default values and ranges in Table~\ref{tab:sgs_params}, including $K_{eddy}=10^3\ \mathrm{m^4\,s^{-1}}$, are those used for the reported 35-member cloud production ensemble. Developmental configurations not used to compute the reported ensemble statistics are not part of the analyzed experiment.

\subsection{Loss Function}
This work aims to optimize the QG low-resolution model by minimizing the discrepancy in the primary dynamical fields in a long 180-day simulation: Potential Vorticity (PV) and Streamfunction ($\psi$). For the high-fidelity simulations, the error metric is computed over the final 30 days of the 180-day integration, when the large initial adjustment has subsided and the flow samples a comparatively settled late-time statistical regime. Because the implementation is freely evolving, we use ``late-time statistical state'' rather than implying a continuously forced stationary climate equilibrium. Our goal is to identify a parameter configuration that makes the low-resolution time-mean fields statistically resemble the high-resolution reference while recognizing that the measured discrepancy includes resolution-dependent dynamics and independently generated native-grid initial conditions.
Because the QG flow is freely evolving and chaotic, early-time mismatch is dominated by spin-up, phase adjustment, and initial-condition realization differences rather than by the persistent statistical effects of the subgrid parameters. Evaluating the settled, later part of the simulation emphasizes time-mean structural statistics and avoids penalizing trajectory-level phase differences that are not meaningful tuning targets.

The implemented loss compares \emph{time-mean fields}; it is not a time average of RMSE values computed from paired instantaneous HR and LR fields. Let $\mathcal{T}$ denote the selected evaluation times. For each resolution $R\in\{HR,LR\}$, the layer-mean PV fields are first computed as
\begin{equation}
    \overline{q_i}^{R}=\frac{1}{|\mathcal{T}_R|}\sum_{t\in\mathcal{T}_R}q_i^R(t),
\end{equation}
after which the corresponding streamfunctions $\overline{\psi_i}^{R}$ are diagnosed by applying the model's spectral PV inversion to $\overline{q_i}^{R}$. The depth-weighted barotropic time means are then
\begin{equation}
    \overline{q}_{bt}^{R}=\frac{H_1\overline{q_1}^{R}+H_2\overline{q_2}^{R}}{H_1+H_2},\qquad
    \overline{\psi}_{bt}^{R}=\frac{H_1\overline{\psi_1}^{R}+H_2\overline{\psi_2}^{R}}{H_1+H_2}.
\end{equation}
Let $\mathcal{C}$ denote the periodic box-filter coarse-graining operator that maps the $512\times256$ HR grid to the $64\times32$ LR grid. For either field $X\in\{q_{bt},\psi_{bt}\}$, the implemented spatial NRMSE is
\begin{equation}
    \operatorname{NRMSE}(X)=
    \frac{\left[\frac{1}{N_{LR}}\sum_{m=1}^{N_{LR}}
    \left(\overline{X}^{LR}_m-\mathcal{C}(\overline{X}^{HR})_m\right)^2\right]^{1/2}}
    {\sigma_m\!\left[\mathcal{C}(\overline{X}^{HR})\right]+10^{-20}},
\end{equation}
where $\sigma_m$ is the spatial standard deviation over the LR grid. The scalar objective is
\begin{equation}
    \mathcal{L}=w_q\operatorname{NRMSE}(q_{bt})+w_{\psi}\operatorname{NRMSE}(\psi_{bt}).
\end{equation}
Thus, temporal averaging precedes PV inversion, barotropic projection, HR coarse-graining, and the final spatial RMSE. No instantaneous HR--LR trajectory pairing enters the loss.

The weights are set to $w_q = 0.6$ and $w_{\psi} = 0.4$. This prioritization reflects the physical nature of the fields: Streamfunction ($\psi$) is an inverse-Laplacian integrated quantity, naturally smoothing out errors and exhibiting slower divergence from the reference. In contrast, Potential Vorticity ($q$) contains high-frequency spatial variability, sharp gradients, and fine-scale filaments that are highly sensitive to parameter changes. By assigning a higher weight to $q$, we force the optimizer to tackle the more difficult challenge of correcting these fine-scale structural errors, rather than simply matching the smoother large-scale flow. We note that this barotropic objective does not explicitly constrain baroclinic structure or spectral fluxes; however, the optimized parameters show independent validation beyond the loss function through representative-case diagnostics in Section~3.3 and ensemble-level analyses in Sections~3.2 and~3.4.

For the low-fidelity stage, the loss is computed over the full 30-day scout simulation using the same weighted formulation. We do not treat this transient-window loss as a substitute for the late-time 180-day objective. Instead, it is used only as a screening signal for learning the broad parameter-to-loss structure and rejecting clearly unfavorable regions before high-fidelity refinement. Parameter-dependent stochastic correction and dissipation affect the model's energy and potential-vorticity adjustment from the beginning of an integration, so the transient response can retain useful information about broad parameter sensitivity even when its absolute loss is not well calibrated to the late-time objective. Consistent with this interpretation, repeating the scout stage with 25-day and 50-day simulations produced the same qualitative screening and optimization conclusions as the 30-day configuration. The final parameter assessment remains based on 180-day simulations evaluated over their final 30 days. The different dynamical regimes sampled by the short and long windows help explain the modest point-wise fidelity correlation ($r \approx 0.37$; Section~3.4), while the full-simulation refinement prevents the transient proxy from determining the final solution.

\subsection{Linear Optimization via Green's Function Optimization}
In Experiment 1, we implemented a simplified one-at-a-time finite-difference screening approach inspired by Green's-function calibration methods in the ocean and Earth-system modeling literature \citep{menemenlis2005using, Strobach2022, Carroll2020}. Full Green's-function formulations assemble model-response sensitivities in a weighted linear inverse problem; our scalar-loss implementation instead uses those local sensitivities to rank parameters and construct a bounded projected update. It is therefore a local sensitivity-gradient probe rather than a reproduction of the full Green's-function inverse formulation. We estimated the sensitivity gradient by perturbing each parameter individually by 50\% around a base state. Because the parameters have different units and span very different numerical ranges, the finite-difference sensitivities were not computed with respect to the raw dimensional parameter values. Instead, each linear parameter was mapped to its bounded coordinate in $[0,1]$, and parameters sampled logarithmically were mapped to a normalized base-10 logarithmic coordinate. The resulting dimensionless-coordinate sensitivities are therefore comparable across parameters and were used both to rank the four active parameters and to construct a bounded projected update (Appendix~A.1). To test whether the screening conclusion depended on the relatively large 50\% perturbation, we repeated the one-at-a-time sensitivity calculation using 5\% and 20\% perturbations. The active-parameter selection and qualitative screening conclusion were unchanged, indicating robustness over the tested perturbation amplitudes. While computationally cheap ($n+1$ simulations for a single perturbation amplitude), this method assumes that one-at-a-time finite differences provide a useful local ranking and that parameter interactions around the base state are not dominant, assumptions examined in our results. For the primary 50\% screening, the method uses 7 full 180-day simulations for 6 parameters: one simulation at the default parameter vector and six simulations in which the parameters are perturbed one at a time.

\subsection{Non-linear optimization using ML}
To establish the capabilities of data-driven methods, we conducted three related experiments. First, we used Random Search (LHS; Experiment 0) as an ``ignorant baseline'' to benchmark the optimization landscape and as ``offline'' training data for ML methods. The baseline LHS design creates $4n$ samples for $n$ parameters; for the full six-parameter experiments, it therefore generates 24 parameter-combination samples. The separate four-parameter hybrid warm start contains 16 mixed-design evaluations and is described in Section~2.5 and Appendix~A.2. We found that fewer than 15 samples are inadequate for ``offline'' ML model training.

For nonlinear ML approaches, we implemented a Gaussian Process ensemble optimizer (Experiment 2). As detailed in Appendix~A.4, the production surrogate combines eight GP models with Mat\'ern and radial-basis-function kernels at different length scales; individual kernels include a white-noise contribution. This ensemble retains the probabilistic uncertainty estimates used for acquisition while reducing dependence on a single smoothness assumption. GP training scales cubically with the number of observations, which can become restrictive as the training set grows. For Exp 2, the ensemble was initially fit with 24 LHS samples (full 180-day simulations); then, in ``online'' training, it proposed the next candidate parameter set and was updated with the resulting loss until reaching a total budget of 50 full 180-day evaluations, inclusive of the initial 24 LHS samples.

We also explored a direct Neural Network optimization (Experiment 3), deployed using all six candidate inputs to learn the parameter-to-loss mapping directly. In this experiment, the iteration process is exactly the same as in Experiment 2, except that we use an MC-dropout NN instead of a GP, with the same total budget of 50 full 180-day evaluations inclusive of the initial 24 LHS samples. Experiment 3 uses only 180-day evaluations, whereas Experiment 6 operates on the four GFO-selected active parameters under the 30-day/180-day schedule (Section 2.5).

\subsection{Multi-Fidelity Hybrid Approach}
\begin{figure}[ht]
    \centering
    \includegraphics[width=\textwidth]{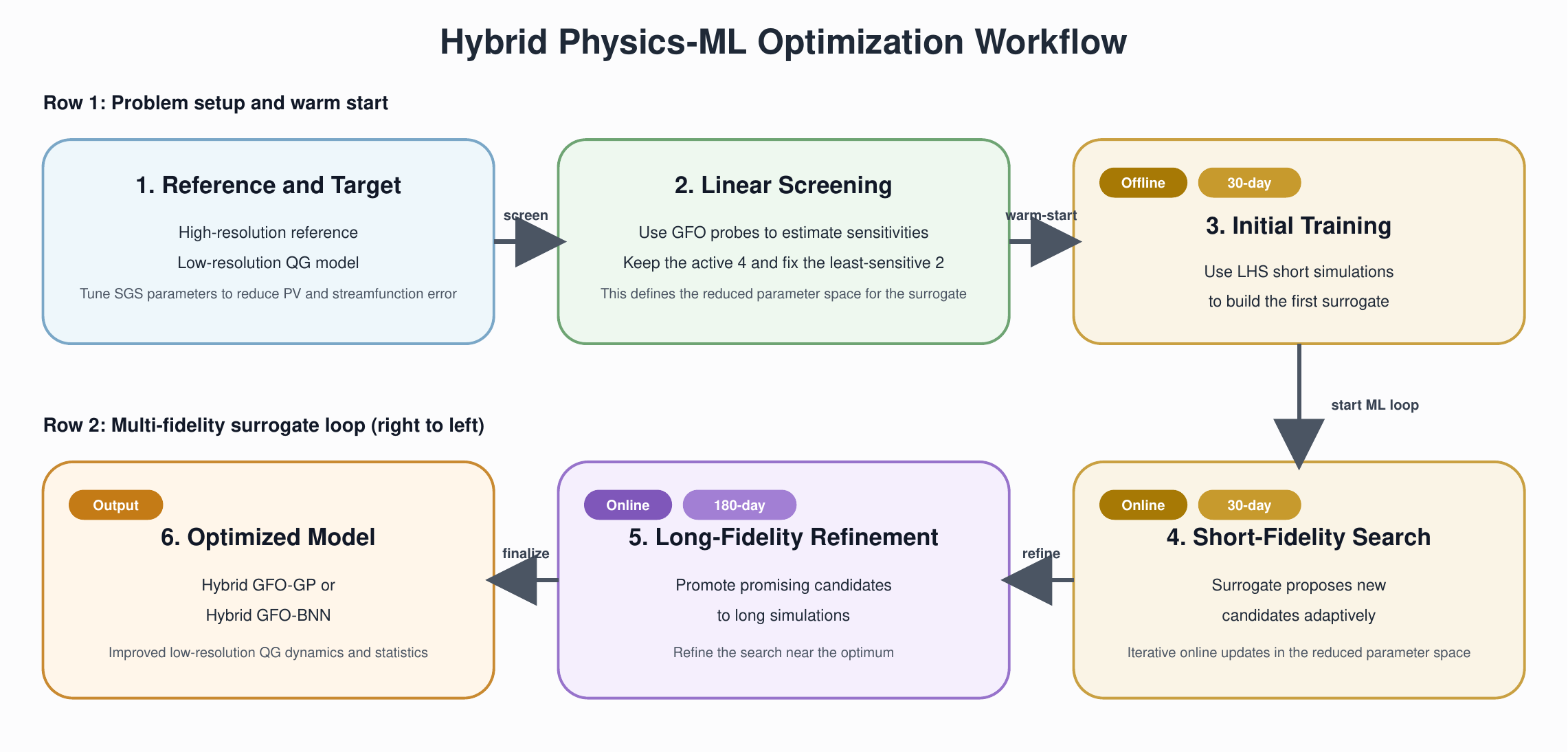}
    \caption{Schematic of the proposed hybrid physics-ML optimization workflow. The top row shows the problem setup, the linear Green's Function Optimization (GFO) screening step used to identify the active parameter subset, and the \textit{offline} surrogate warm start using short 30-day simulations. The bottom row shows the right-to-left multi-fidelity optimization loop: \textit{online} short-fidelity exploration, \textit{online} long-fidelity refinement with 180-day simulations, and the final optimized low-resolution model. The resulting GFO-MultiGP and GFO-MultiNN frameworks focus expensive evaluations only on promising regions of parameter space.}
    \label{fig:hybrid_workflow}
\end{figure}

The core contribution of this work is the Hybrid Multi-Fidelity Strategy (Experiments 5 \& 6), designed to overcome the ``cold start'' problem of Bayesian methods and the linearity limit of GFO. We also introduce a comparison experiment (Experiment 4, MultiNN) that applies the multi-fidelity NN pipeline to all six candidate inputs without prior GFO screening. Here, ``six inputs'' denotes dimensions varied by the optimizer: only five affect the archived dynamics because $C_s$ is an inactive null control. Thus Experiment 4 tests the practical value of screening a candidate list containing an inactive control, not a comparison in which all six dimensions are dynamically effective. The protocol proceeds in three steps. First, we utilize linear GFO probes (taken from Exp 1) to perform sensitivity-based screening (related to, but simpler than, the gradient-covariance active subspace framework of \citealp{Constantine2015}), ranking parameters by the absolute magnitudes of their normalized-coordinate sensitivities. This screening is carried out separately for each ensemble member, because the system state differs across the ensemble. Second, via Subspace Reduction, the optimizer selects the four highest-ranked parameters, fixes the two remaining parameters to the bounded values suggested by that member's GFO step, and restricts the subsequent search to the four-dimensional active subspace. The choice is a fixed top-four ranking rule, not a cumulative-variance or 90\% threshold. Finally, in the Multi-Fidelity Refinement stage, we hand off the reduced-dimension problem to a non-linear surrogate optimizer (GP or NN). The optimizer first explores using 30-day simulations and then transitions to 180-day simulations for refinement. The initial 16 short-run evaluations comprise the default reference, seven Latin-hypercube points, and eight scrambled Sobol points, followed by eight acquisition-guided short-run evaluations. From iteration 25 onward, the workflow uses full 180-day evaluations, with a maximum of 50 surrogate iterations. At the fidelity transition, the current best candidate and default reference are re-evaluated at 180 days to anchor the surrogate to the high-fidelity objective. Figure~\ref{fig:hybrid_workflow} summarizes this workflow.

\begin{table}[H]
    \centering
    \caption{Summary of experiments}
    \label{tab:experiments}
    \begin{tabular}{lccc}
        \hline
        \textbf{Experiment} & \textbf{Methodology} & \textbf{Surrogate} & \textbf{Fidelity} \\
        \hline
        \textbf{Exp 0} & Random Search & None & Single/Multi \\
        \textbf{Exp 1} & GFO & Gradient Projection & Single (180d) \\
        \textbf{Exp 2} & Bayesian Optimization & GP Ensemble & Single (180d) \\
        \textbf{Exp 3} & Deep Learning & MC-Dropout NN & Single (180d) \\
        \textbf{Exp 4} & MultiNN & Neural Network & Multi-Fidelity (30d/180d)\\
        \textbf{Exp 5} & GFO-MultiGP & Probabilistic GP & Multi-Fidelity (30d/180d)\\
        \textbf{Exp 6} & GFO-MultiNN & Neural Network & Multi-Fidelity (30d/180d)\\
        \hline
    \end{tabular}
\end{table}

To assess all methods under a common benchmark, we performed a 35-member ensemble experiment. Each ensemble member was generated by varying the random phases and amplitude used to initialize the native-grid PV fields. This amplitude controls the strength of the initial vertically sheared, or baroclinic-like, component; it is an initial-condition ensemble factor rather than an additional continuous forcing operator omitted from the equations above. Each realization therefore samples a different initial flow state and adjustment pathway. The same 35 ensemble configurations were used across all methods to enable paired comparisons in the Results section.

\section{Results}

We evaluated seven distinct optimization strategies (Table \ref{tab:experiments}) to compare linearity, surrogate choice, dimensionality reduction, and multi-fidelity scheduling within several end-to-end pipelines. Experiment 0 used Random Search (LHS) to establish an ``ignorant'' sampling baseline. Experiment 1 used the linear Green's Function Optimization approach to establish a local sensitivity-screening baseline. Experiments 2 and 3 benchmarked standard single-fidelity black-box optimization using Gaussian Processes (GP) and Direct Neural Networks, respectively, operating on six candidate inputs, of which five were dynamically effective. Experiments 5 (GFO-MultiGP) and 6 (GFO-MultiNN) tested the proposed multi-fidelity framework, which integrates linear subspace screening with non-linear surrogates (GP or NN) operating on the four active parameters identified by GFO screening. Experiment 4 (MultiNN) provides the corresponding unscreened six-input multi-fidelity comparison. Since Experiments 4 and 6 differ simultaneously in screening, search dimension, fixed parameter values, and initialization, their comparison evaluates the complete pipeline change and does not isolate the effect of any one component.

To compare computational cost across methods on a common footing, we define method efficiency as the total simulated days required to reach 95\% of a method's maximum achieved improvement. The denominator in the reported ``simulation-days per 1\% improvement'' is therefore the improvement achieved at this 95\%-of-best threshold, rather than the unconstrained final asymptote. This threshold marks the onset of practical saturation and avoids over-penalizing methods for marginal late-stage gains. Separate post-hoc robustness experiments---including alternative scout durations, alternative GFO perturbation amplitudes, and matched-candidate promotion diagnostics---are validation analyses and are not included in the primary optimization-cost curves.

\subsection{Linear Optimization: Green's Function Optimization}

\begin{figure}[htbp]
    \centering
    \includegraphics[width=.8\textwidth]{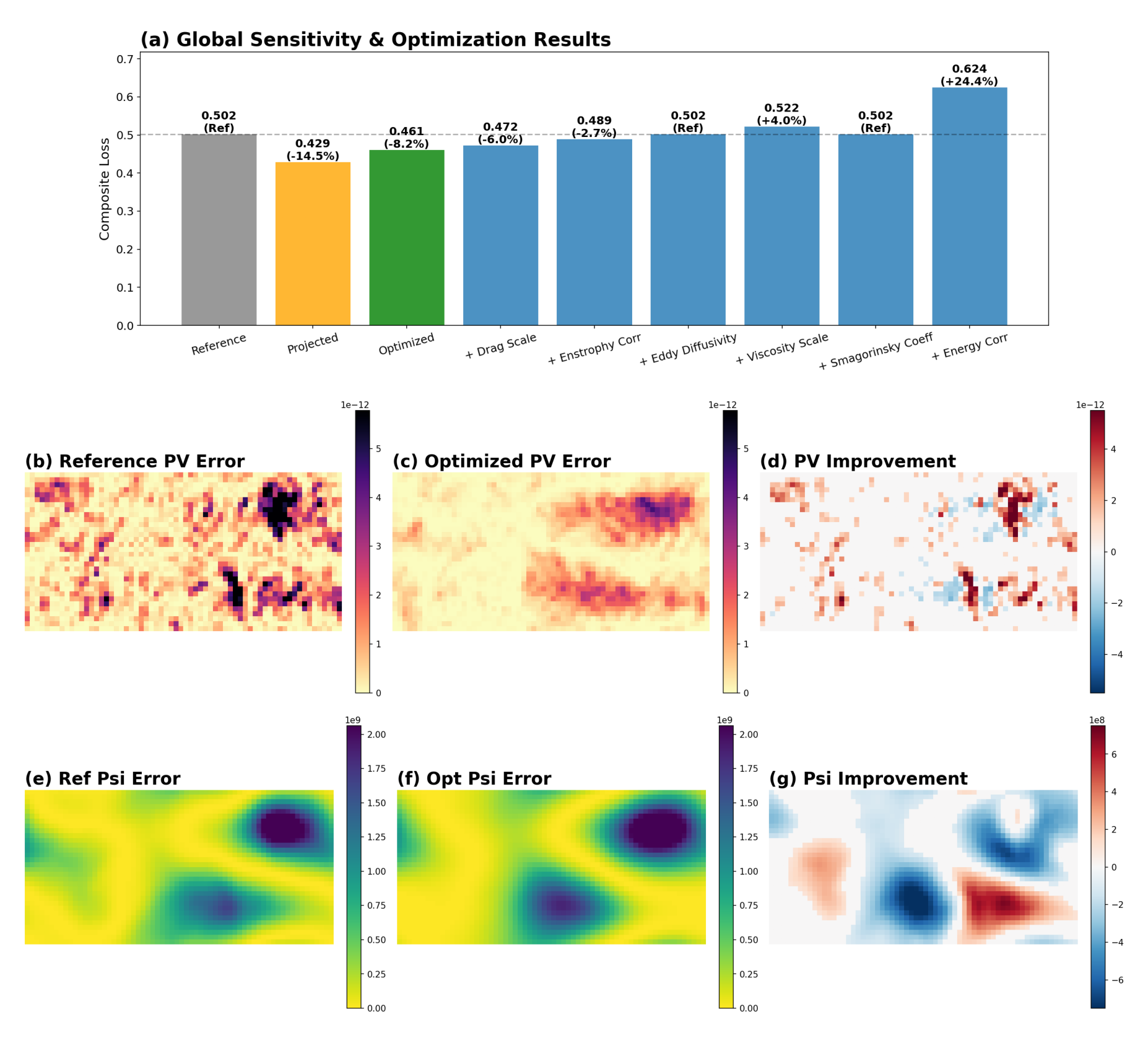}
    \vspace{-.5cm}
    \caption{Green's Function Optimization Diagnostic Analysis. (a) Global Cost Comparison shows the Reference (Gray), Projected (Blue), and Actual Optimized (Green) performance, alongside individual parameter sensitivities. (b-c) Potential Vorticity (PV) error maps before and after GFO optimization. (d) PV closeness map, computed from the difference between the absolute errors, where red indicates that GFO optimization reduced the error (improved the model) and blue indicates that it increased the error (made the model worse). (e-f) Streamfunction ($\psi$) error maps before and after GFO optimization. (g) Streamfunction closeness map defined in the same way, with red denoting improvement and blue denoting degradation.}

    \label{fig:gf_diagnostics}
\end{figure}

The linear Green's Function Optimization (GFO) yielded an immediate performance gain. As shown in Figure \ref{fig:gf_diagnostics} for one ensemble case, seven screening runs (one default plus six one-at-a-time perturbations) produced a projected parameter vector whose independent 180-day validation reduced global error by 8.2\% ($0.502 \to 0.461$). We include that validation in the cost-to-achieve metric: the validated GFO result therefore requires $8\times180=1{,}440$ simulation-days, corresponding to 175.6 simulation-days per 1\% improvement.

The diagnostic response analysis (Figure \ref{fig:gf_diagnostics}a) identified Energy Correction ($\epsilon_{back}$; plotted response score $+2.52$) and Linear Drag ($\alpha_r$; plotted response score $-0.73$) as the two strongest controls for this representative member. These plotted scores follow the diagnostic response convention used in the figure and should not be read as the signed normalized-coordinate derivative $J_i^{(z)}$ in Appendix~A.1. The bounded GFO update itself uses $J_i^{(z)}$ and moved the configuration toward weaker backscatter magnitude and stronger drag. This rebalancing was associated with a more stable large-scale energy balance and smaller field biases.

Critically, these results define the role of linear methods in the tuning pipeline. While the GFO approach correctly identified the \textit{direction} of the necessary parameter shifts, it failed to predict the \textit{magnitude} of the improvement due to non-linear saturation. The linear projection (Supplementary Figure S1) was ``over-optimistic,'' predicting a cost drop to 0.429 (14.5\% improvement), whereas the actual system saturated at 0.461 (8.2\% improvement). This discrepancy demonstrates that the local sensitivity gradient is a powerful estimator for \textit{screening}, because it cheaply ranks parameters and prunes poor regions, but validating the precise optimum requires a non-linear optimizer capable of navigating the diminishing returns of the loss landscape.

\subsection{Nonlinear and Hybrid Optimization Strategy}

\begin{figure}[ht]
        \centering
        \includegraphics[width=.95\textwidth]{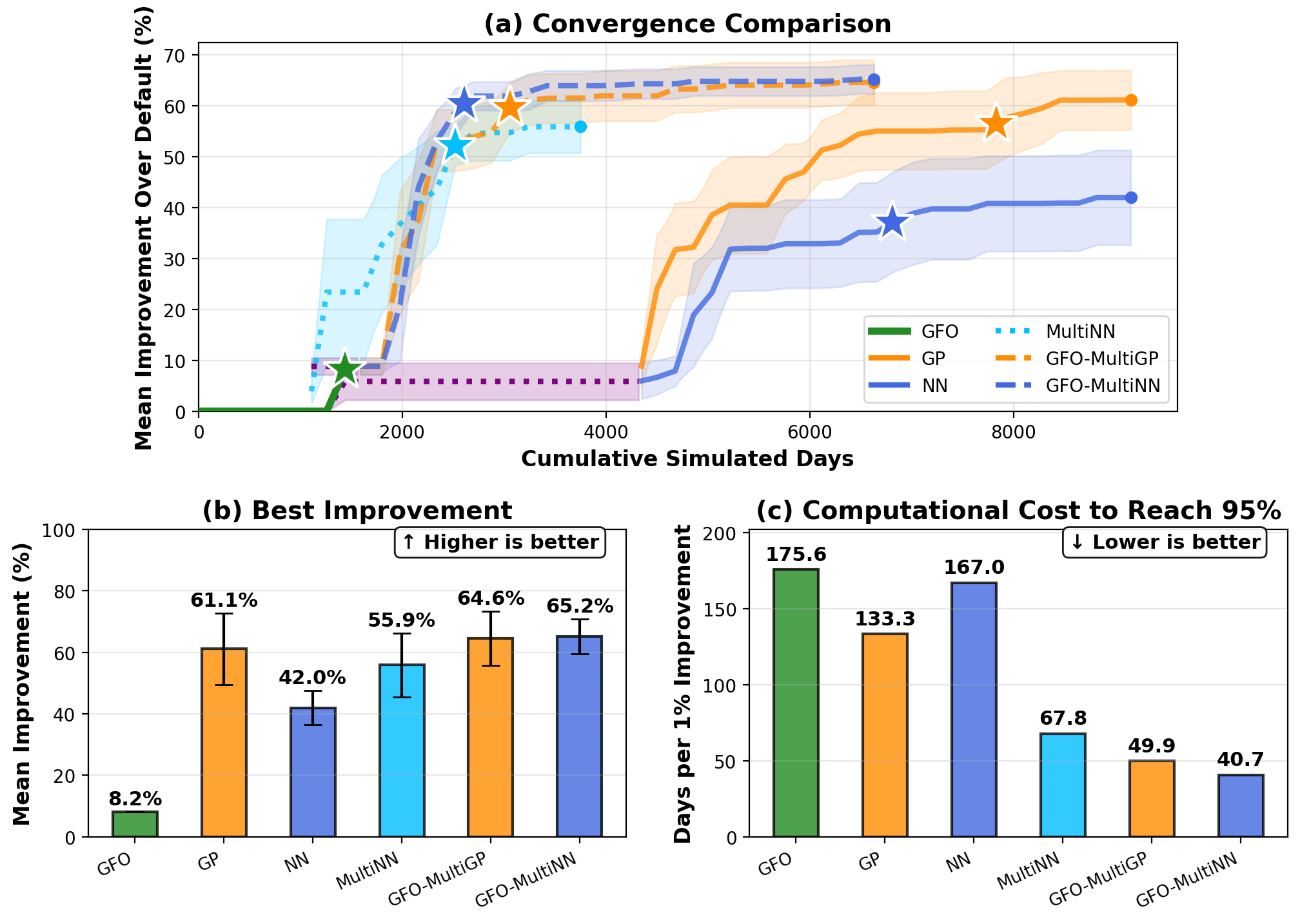}
        \caption{Ensemble Optimization Comparison ($N=35$ ensembles for each method). (a) Convergence time series showing improvement over default for GP, NN, MultiNN, and the Hybrid methods (GFO-MultiGP and GFO-MultiNN); the standalone GFO point is placed after seven screening runs and its independent validation run. Stars mark points reaching within 5\% of each method's best improvement. (b) Mean best improvement achieved by each method with standard deviation error bars. (c) Computational cost measured as simulation days per 1\% improvement at the 95\% threshold; standalone GFO includes its additional 180-day validation run. The hybrid methods achieve significantly better efficiency ($p < 0.01$, Wilcoxon signed-rank test) with comparable or superior final improvement.}
        \label{fig:ensemble_compare}
    \end{figure}

Using this common ensemble benchmark, Figure \ref{fig:ensemble_compare} compares the performance of the Hybrid Multi-Fidelity approach against the single-fidelity methods. Experiments 2 and 3 were each allocated a budget of 50 high-fidelity (180-day) evaluations, totaling 9,000 simulation-days per method. Experiment 1 uses seven GFO screening runs plus the independent validation run required to establish its achieved improvement. The hybrid methods reuse the seven screening runs but do not require the standalone GFO validation as an algorithmic input; their subsequent candidates are evaluated within their own multi-fidelity trajectories. Each hybrid then allocates a maximum of 50 surrogate-guided iterations: the first 24 iterations ($6k$, where $k=4$ active parameters) use cheap 30-day simulations for exploration, and the remaining high-fidelity phase uses 180-day anchoring and refinement evaluations. At the transition, the current best candidate and default baseline are re-evaluated at 180~days to anchor the high-fidelity objective. In practice, all methods reach practical saturation before exhausting their nominal budgets. Under the 95\%-saturation metric, GFO-MultiNN reaches 95\% of its maximum achieved improvement after 2,520 simulation-days, while GFO-MultiGP reaches the same threshold after 3,060 simulation-days. The corresponding saturation point occurs after 7,740 simulation-days for the standalone GP and after 6,660 simulation-days for the standalone NN. Thus, the hybrid methods reach near-maximum performance with roughly 54--67\% less simulation time than the standalone ML approaches while also attaining higher final improvements (64.6\% and 65.2\% vs.\ 61.1\% and 42.0\%; Figure~\ref{fig:ensemble_compare}b). Statistical significance was assessed using a two-sided Wilcoxon signed-rank test on the paired ensemble improvements. Because the hybrid and standalone methods differ in more than one design choice, these gains should be interpreted as end-to-end pipeline gains rather than as isolated contributions from fidelity scheduling alone.

To benchmark performance, we first evaluated the single-fidelity learning approaches. The Random Search baseline (Experiment 0; LHS) provided minimal improvement, serving primarily to bound the difficulty of the landscape. The Direct Neural Network (Experiment 3) demonstrated learning capability, achieving a maximum improvement of 42.0\%, but required 6,660 simulation-days to reach 95\% saturation, corresponding to a poor efficiency of 167.0 simulation-days per 1\% improvement. The standalone GP ensemble (Experiment 2) proved more robust, achieving a respectable 61.1\% improvement, but it still required 7,740 simulation-days to reach 95\% saturation, yielding an efficiency of 133.3 simulation-days per 1\% gain.

As an end-to-end comparison, we tested the six-input multi-fidelity strategy in Experiment 4 (MultiNN); five inputs were dynamically effective and $C_s$ served as a null control. This setup achieved a 55.9\% improvement at a computational saturation limit of 3,600 simulation-days (67.8 simulation-days per 1\% gain), but it fell short of the tested GFO-MultiNN pipeline in both maximum accuracy and cost-efficiency. Because the experiments differ in several design choices, including removal of the null dimension, this result supports an advantage for the complete screened pipeline but does not isolate which component produces the gain.

The proposed Hybrid Multi-Fidelity Strategy changed this trajectory. Both hybrid variants leveraged the initial GFO screening (Fig. \ref{fig:ensemble_compare}a, green line) to bypass the ``cold start'' phase. Experiment 5 (GFO-MultiGP) rapidly converged to a 64.6\% improvement, while Experiment 6 (GFO-MultiNN) achieved the highest overall performance with a 65.2\% mean improvement. By utilizing 30-day ``scout'' simulations to filter the search space, these methods reserved expensive 180-day simulations only for high-probability candidates.

Figure \ref{fig:ensemble_compare} provides a comprehensive comparison of these strategies. The impact of the hybrid architecture is most evident in the computational cost (Fig. \ref{fig:ensemble_compare}c). The GFO-MultiGP and GFO-MultiNN reduce the cost to 49.9 and 40.7 simulation-days per 1\% improvement, respectively. Within the set of methods tested here, GFO-MultiNN delivered the best combination of efficiency and final improvement, representing a 3.3x efficiency gain over the standalone GP ensemble ($p < 0.01$, Wilcoxon signed-rank test) and a 4.1x gain over the pure NN, while also exceeding their final accuracy (65.2\% vs.\ 61.1\% and 42.0\%). Taken together, these results support the view that physics-informed screening combined with non-linear multi-fidelity refinement can be a particularly effective pathway for this QG tuning problem.

\subsection{Optimized QG Turbulence Model}

To illustrate the case-level behavior behind the ensemble statistics, we discuss one randomly selected ensemble member from the 35-member benchmark described in Section~2.5. For this representative case, the optimized model achieves a final loss of 0.129, effectively matching the best-performing standalone GP benchmark (0.130) and producing a substantially lower loss than linear GFO (0.461). This case illustrates that the hybrid method can retain much of the non-linear search capability while using fewer full-fidelity evaluations; statistical comparisons are based on the paired 35-member ensemble rather than this individual example. The optimized parameterization reduced the Root Mean Square Error (RMSE) of the depth-weighted streamfunction diagnostic by 81\% ($1.88 \times 10^{4} \to 3.54 \times 10^{3}\ m^2/s$) and the corresponding PV diagnostic by 57\% ($4.13 \times 10^{-7} \to 1.76 \times 10^{-7}\ s^{-1}$) compared with the default low-resolution configuration. Table~\ref{tab:opt_params} reports the corresponding optimized parameter values for this representative member.

\begin{table}[H]
    \centering
    \caption{Optimized Subgrid Parameters for one randomly selected ensemble member: Comparison of Default (baseline), Linear GFO (Exp 1), and GFO-MultiNN (Exp 6) values. Percentage changes ($\Delta\%$) relative to the Default configuration are shown in parentheses; for the negative Energy Correction coefficient, positive percentages denote reductions in absolute magnitude toward zero.}
    \label{tab:opt_params}
    \begin{tabular}{lccc}
        \hline
        \textbf{Parameter (Symbol)} & \textbf{Default} & \textbf{GFO Opt ($\Delta\%$)} & \textbf{GFO-MultiNN ($\Delta\%$)} \\
        \hline
        Viscosity Scale ($\alpha_{\nu}$) & 0.50 & 0.51 (+2\%) & 0.99 (+98\%) \\
        Drag Scale ($\alpha_{r}$) & 0.50 & 0.52 (+4\%) & 1.11 (+122\%) \\
        Eddy Diffusivity ($K_{eddy}$) & 1000 & 1000 (0\%) & 1000 (Fixed) \\
        Smagorinsky Coeff ($C_s$) & 0.015 & 0.015 (0\%) & 0.015 (Fixed) \\
        Energy Correction ($\epsilon_{back}$) & -2.0e-3 & -1.5e-3 (+25\%) & -2.8e-4 (+86\%) \\
        Enstrophy Correction ($\epsilon_{ens}$) & 3.0e-9 & 2.8e-9 (-6\%) & 6.8e-9 (+127\%) \\
        \hline
    \end{tabular}
    \vspace{0.2cm}
\end{table}

    \begin{figure}[ht]
        \centering
        \includegraphics[width=0.9\textwidth]{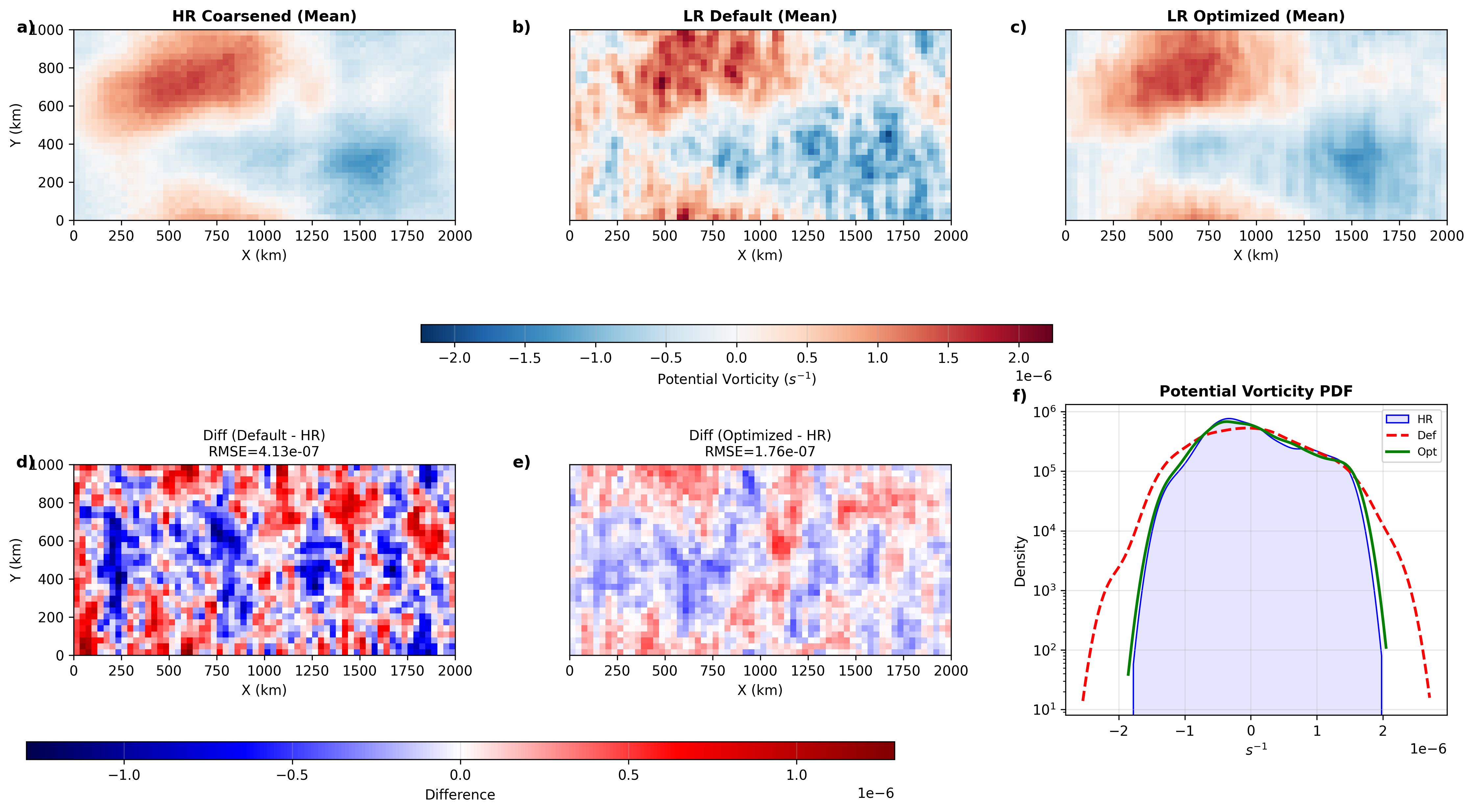}
        \caption{Potential Vorticity ($q$) Field Comparison. The optimization improves the representation of fine-scale PV filaments and gradients, crucial for accurate enstrophy dynamics.}
        \label{fig:field_q}
    \end{figure}
    
These statistical improvements are confirmed by the field reconstructions shown in Supplementary Figure S2. The optimized model effectively corrects the zonal jet structure and reduces large-scale biases. Similarly, Figure \ref{fig:field_q} demonstrates that the optimization improves the representation of fine-scale PV filaments and gradients, crucial for accurate enstrophy dynamics.

    \begin{figure}[ht]
        \centering
        \includegraphics[width=.80\textwidth]{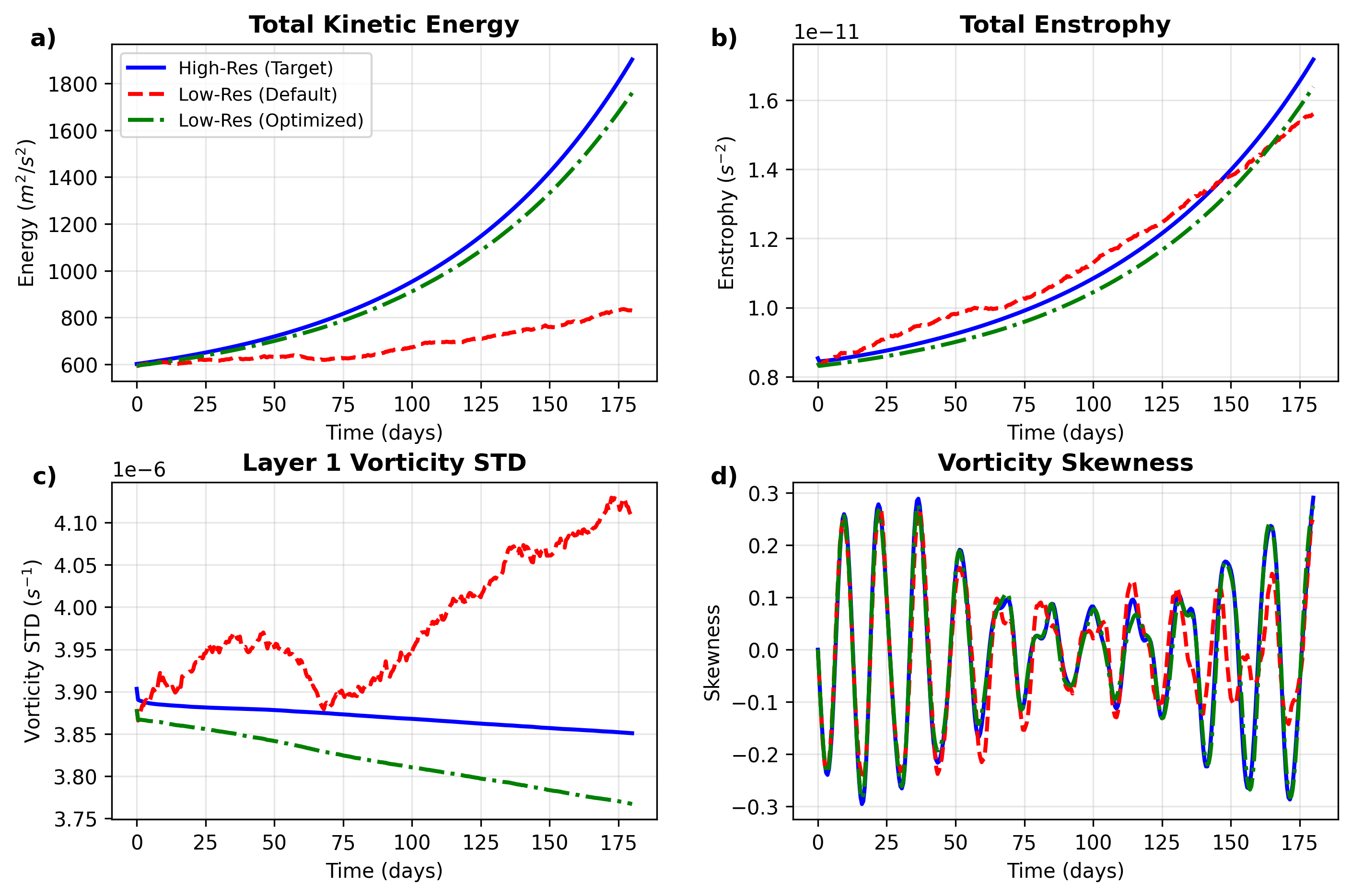}
        \caption{Three-Way Dynamics Comparison. (a) Total Kinetic Energy and (b) Total Enstrophy time series show the optimized low-resolution model (green) tracking the high-resolution reference (blue) much more closely than the default baseline (red). (c) Layer 1 Vorticity standard deviation shows that the optimized run suppresses the spurious variance growth seen in the default case and remains much closer to the high-resolution reference. (d) Vorticity skewness shows that the optimized run also better reproduces the time-varying asymmetry of the reference vorticity distribution.}
        \label{fig:dynamics_3way}
    \end{figure}

\subsection{Multi-Fidelity Analysis: How the Surrogate Learns}

The method also improves the bulk energy and enstrophy statistics of the simulation. As shown in Figure \ref{fig:dynamics_3way}, the optimized model follows the Total Kinetic Energy and Total Enstrophy evolution of the high-fidelity reference more closely and substantially reduces the drift observed in the default baseline. The lower panels further show that the optimized parameters keep the Layer 1 vorticity variance closer to the reference trajectory and better reproduce the evolving skewness of the vorticity distribution, indicating improvement in representative higher-order dynamical diagnostics beyond the bulk integral measures.

\begin{figure}[htbp]
    \centering
    \includegraphics[width=1.0\textwidth]{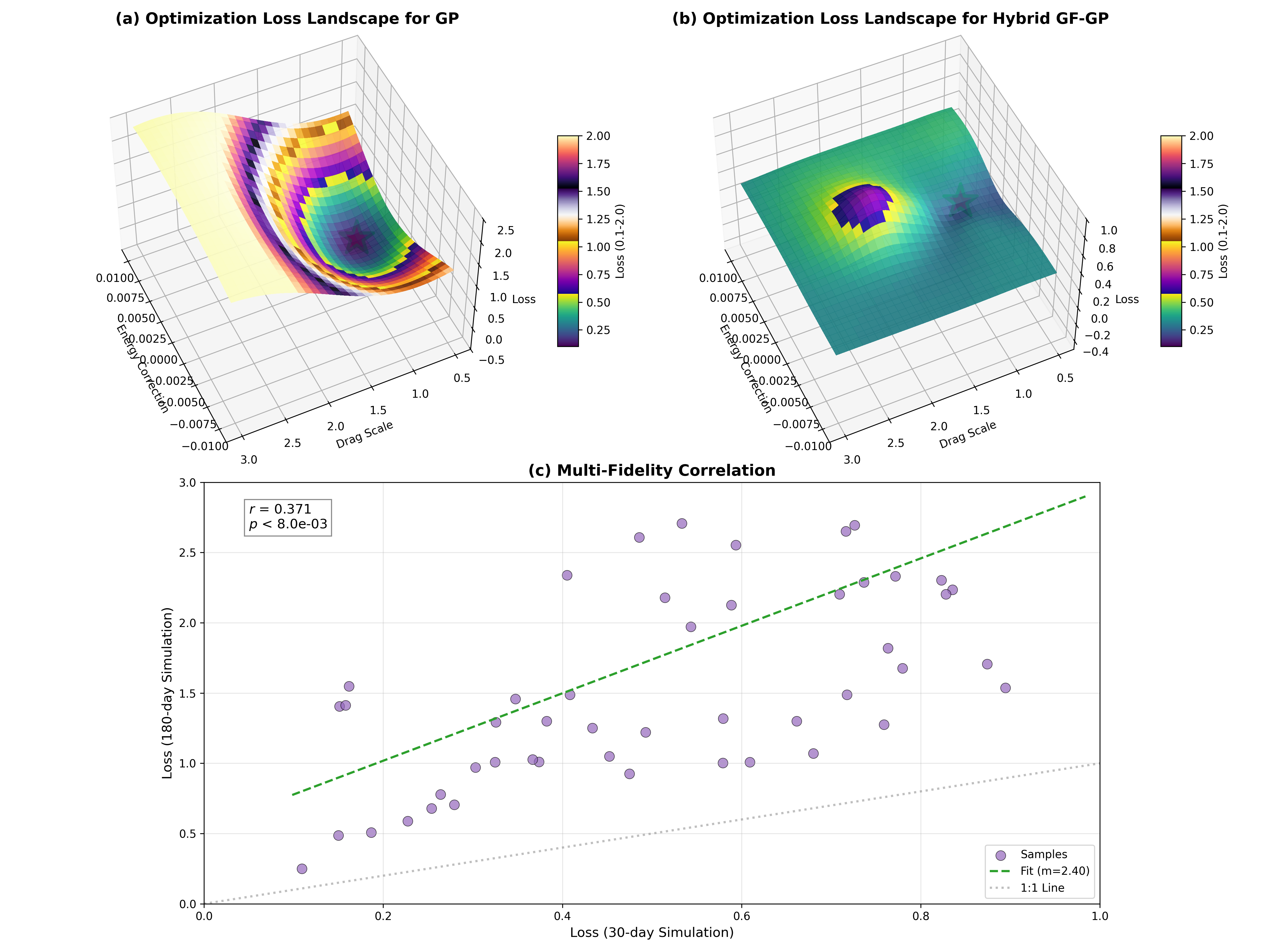}
    \caption{Multi-Fidelity Optimization Landscape. (a) High-fidelity (180-day) 3D loss landscape from the standalone GP ensemble, showing the reference global structure. (b) Low-fidelity (30-day) 3D loss landscape from the GFO-MultiGP exploration stage, showing a broadly similar low-loss region. (c) Fidelity correlation indicating that while the point-wise correlation is weak ($r \approx 0.37$), the positive trend and qualitative basin agreement allow the low-fidelity model to serve as an effective negative filter.}
    \label{fig:loss_landscape}
\end{figure}

A critical insight from our experiments is that the low-fidelity stage is more useful as a screening proxy than as a quantitative predictor of 180-day loss. As shown in Figure~\ref{fig:loss_landscape}(c), the Pearson correlation between the 30-day and 180-day loss values is modest ($r \approx 0.37$), indicating substantial point-wise differences in absolute loss magnitude. This is physically expected because the 30-day simulations primarily sample transient adjustment and spin-up, whereas the high-fidelity metric samples the settled final 30 days of a 180-day integration. Nevertheless, parameter changes influence stochastic correction, dissipation, and the developing energy and enstrophy balances during spin-up. The transient response can therefore preserve the broad geometry of favorable and unfavorable parameter regions without reproducing the late-time loss magnitude. The Spearman rank correlation is somewhat higher ($\rho \approx 0.43$, $p < 0.01$), consistent with the ordering of broad parameter quality being more transferable than the raw values.

The surrogate uses this short-run information to map the coarse loss-landscape structure and prioritize candidates, rather than to declare a final optimum from transient behavior. At the fidelity transition, the default and best short-run candidate are re-evaluated with 180-day simulations, after which full-length evaluations refine the search against the late-time objective. The 30-day stage can thus be useful even with modest cross-fidelity correlation because errors in its absolute calibration are corrected during high-fidelity refinement. As a robustness check on the scout duration, we repeated the low-fidelity stage with 25-day and 50-day simulations. Both alternatives led to the same qualitative screening and final optimization conclusions as the 30-day setup, indicating that the reported result is not specific to a single short-run duration.

\begin{figure}[ht]
    \centering
    \includegraphics[width=1.0\textwidth]{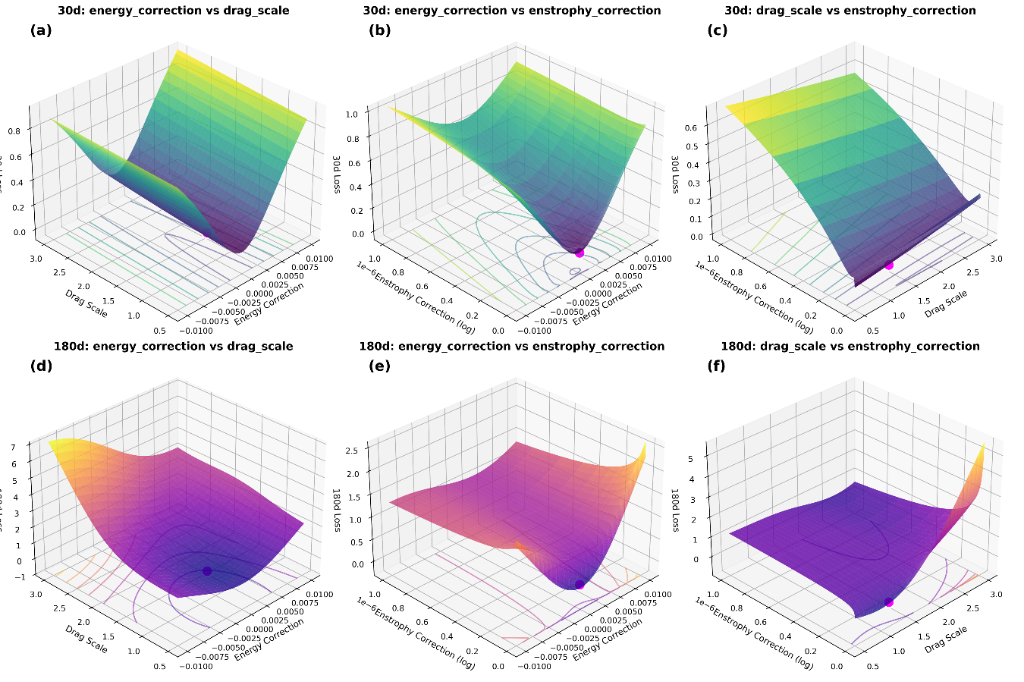}
    \caption{Pairwise Loss Landscape Consistency. Comparison of 30-day (Top Row: a, b, c) and 180-day (Bottom Row: d, e, f) loss landscapes across three different parameter pairings. Despite differences in absolute loss magnitude (z-axis), the low-loss basins occur in qualitatively similar regions of parameter space between fidelities.}
    \label{fig:landscape_pairwise}
\end{figure}

Despite this noise, Figures~\ref{fig:loss_landscape}(a) and~\ref{fig:loss_landscape}(b) still provide a useful qualitative comparison: the broad ``valley'' of low-loss solutions (dark blue regions) appears in the same quadrant of parameter space for both the high-fidelity reference landscape and the low-fidelity hybrid exploration landscape. Because this comparison also changes the optimizer setup and reduced search subspace, we do not interpret Figure~\ref{fig:loss_landscape} alone as a controlled fidelity-only test. Rather, it suggests that the hybrid exploration is being guided toward the same broad basin identified by the high-fidelity objective. A stricter diagnostic comes from the matched-candidate promotion analysis associated with Figure~\ref{fig:landscape_pairwise}, where the same 24 scout candidates are ranked at 30~days and then re-evaluated at 180~days within each ensemble member. We quantify this screening utility using a \textit{promotion success rate}, defined as the fraction of parameter sets ranked in the top quartile by 30-day loss (i.e., the 6 lowest-loss candidates out of the 24 scout simulations) that, when re-evaluated at 180~days, achieve a loss below the 180-day default baseline. Across the 35 ensemble members, the mean promotion success rate is 76\% ($\pm 8\%$, $1\sigma$ across ensemble), corresponding to an approximate 95\% confidence interval of 73.3--78.7\% for the ensemble-mean success rate. In other words, parameter sets that looked promising at 30~days were confirmed as genuine improvements at 180~days roughly three out of four times. We emphasize, however, that this is still a workflow-specific screening diagnostic rather than a complete fidelity-transfer validation: we do not include random-promotion or simple threshold baselines here, and the two fidelities are evaluated over different windows (Section~2.2). Accordingly, we interpret the 30-day stage as a useful \textit{negative filter} within the present pipeline, not as a stand-alone surrogate for the 180-day objective.

Importantly, this qualitative basin agreement is not limited to a single parameter pair. Figure~\ref{fig:landscape_pairwise} shows three representative pairwise sections of the four-parameter active subspace: Energy Correction vs.\ Drag Scale (a,d), Energy Correction vs.\ Enstrophy Correction (b,e), and Drag Scale vs.\ Enstrophy Correction (c,f). In each displayed section, the 30-day simulations tend to identify low-loss regions in the same broad neighborhood as the 180-day reference. These selected comparisons suggest that the low-fidelity model captures useful dominant sensitivities for screening, even when it does not predict the exact loss magnitude; they do not constitute an exhaustive visualization of all six possible active-parameter pairs.

Concretely, the multi-fidelity strategy operates as a two-phase evaluation schedule rather than a coupled discrepancy model. In the \textit{exploration phase} (first $6k = 24$ iterations for $k=4$ active parameters), the optimizer evaluates candidate parameter sets using cheap 30-day simulations. The first $4k = 16$ evaluations form a mixed space-filling warm start consisting of the default reference, seven Latin-hypercube points, and eight scrambled Sobol points (Appendix~A.2). The remaining $2k = 8$ exploration iterations use the surrogate's acquisition function, a weighted combination of Expected Improvement (EI, weight $w_{EI} = 0.6$) and Upper Confidence Bound (UCB, $\kappa = 2.0$) with local penalization to prevent redundant sampling, to propose new candidates (see Appendix~A.5 for the full acquisition formulation). Thompson sampling is used with 10\% probability per iteration for additional exploration. Because the 30-day topology retains useful information about the broad basin of the 180-day objective (Figures~\ref{fig:loss_landscape}--\ref{fig:landscape_pairwise}), this phase identifies promising regions at low cost. In the \textit{refinement phase} (iterations 25+), the evaluation switches to full 180-day simulations. The current best parameters and the default baseline are both re-evaluated at 180~days to establish anchored references, and the surrogate is retrained on the accumulated dataset. The two hybrid variants differ primarily in the surrogate architecture: Experiment 5 (GFO-MultiGP) uses a weighted ensemble of 8 Gaussian Processes with diverse kernels (Mat\'ern 1.5/2.5, RBF at varying length scales), while Experiment 6 (GFO-MultiNN) uses a Neural Network with MC~Dropout \citep{Gal2016} for uncertainty-aware predictions (see Appendix~A for full architectural details).

\subsection{Algorithm Robustness and Parameter Equifinality}

\begin{figure}[ht]
    \centering
    \includegraphics[width=1.0\textwidth]{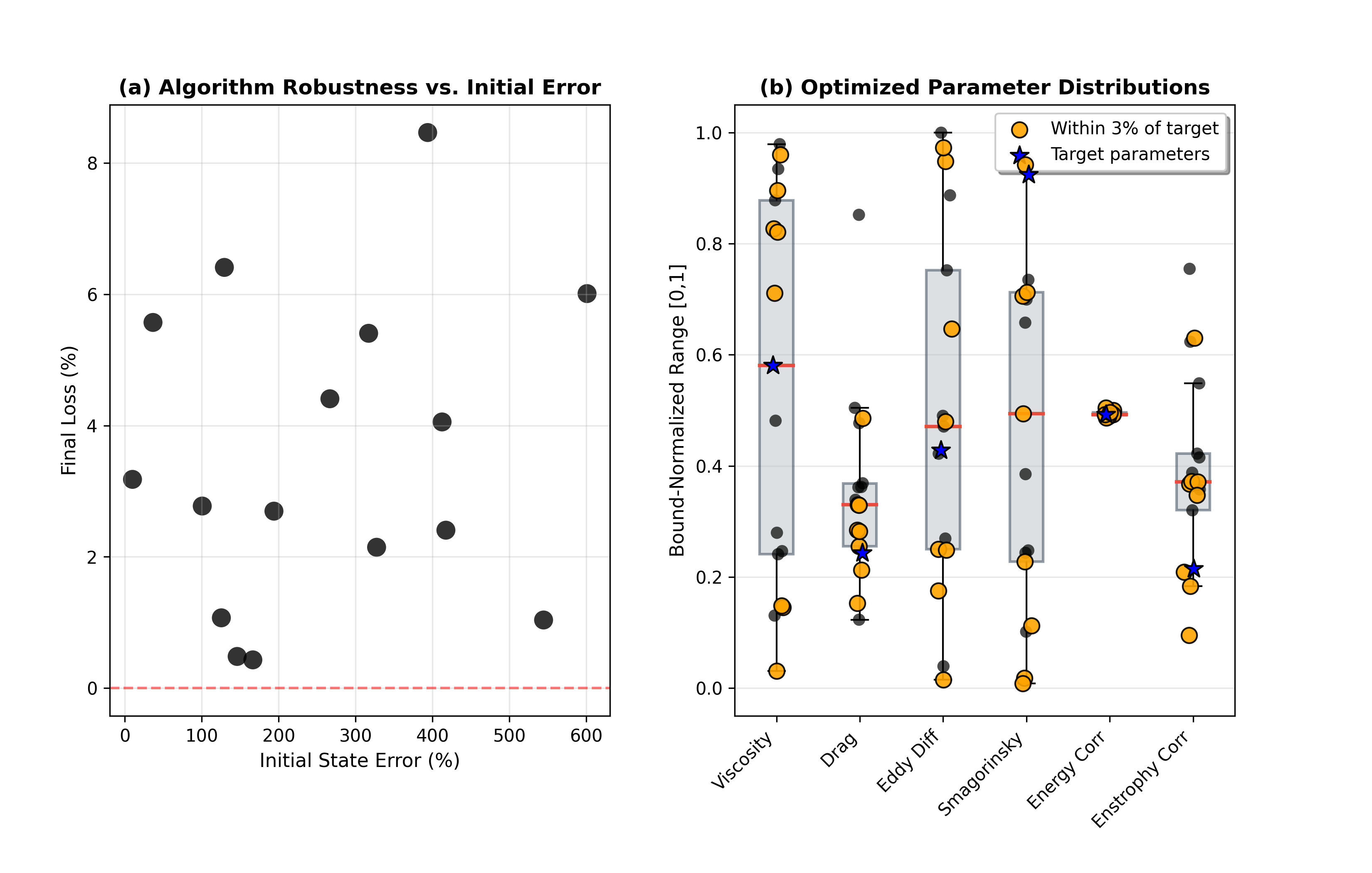}
    \vspace{-1cm}
    \caption{Ensemble validation of algorithmic robustness and parameter equifinality across 17 randomized initializations. (a) Final optimization loss evaluated against initial state error. All runs converge within 9\% of the known-reference target loss despite initial baseline errors exceeding 600\%. (b) Distribution of the optimized subgrid parameters. The best achieved configuration is denoted by blue stars. Parameter sets within 3\% of the best achieved loss (orange markers) exhibit broad variation across implemented parameters, indicating a low-loss equifinal region. Variation in the inactive Smagorinsky coefficient is a null-control result and is not interpreted as physical compensation.}
    \label{fig:equifinality}
\end{figure}

To evaluate the robustness of the hybrid optimization framework and the uniqueness of low-loss parameter solutions, a 17-member identical-twin experiment was conducted. In this setup, the target objective was generated by the low-resolution model driven by a known target parameter set rather than by the standard high-resolution reference. The experiment asks whether the optimizer recovers the hidden parameters and whether exact parameter recovery is necessary to reproduce the target model statistics. Across all 17 members, the model configuration, physical grid, and dynamical initial conditions were identical; only the starting parameter guess was varied, producing initial baseline errors up to 600\% above the target loss (Figure~\ref{fig:equifinality}a).

The Multi-fidelity GFO-BNN framework was robust to these initial guesses: all runs converged within 9\% of the known-reference target loss. The resulting parameter distributions also show a low-loss equifinal region (Figure~\ref{fig:equifinality}b). The energy correction coefficient concentrated near a common value, whereas several implemented controls, including viscosity, drag, and eddy diffusivity, retained broader distributions among solutions within 3\% of the best achieved loss. This indicates that exact recovery of the generating parameter vector was not required to reproduce the target diagnostics within the tested tolerance. The Smagorinsky coefficient also spans a broad range, but because it is inactive in the archived time-stepping implementation, that spread is expected non-identifiability and is excluded from the physical equifinality interpretation. The remaining spread is consistent with compensation among active dissipation and correction terms, although the experiment does not establish that every low-loss parameter difference is dynamically interchangeable.

\section{Discussion: Scalability \& Limitations}

To examine the controls associated with model error, we compared the GFO screening and final GFO-MultiNN adjustments (Table~\ref{tab:opt_params}) for the representative member in Section~3.3. Under the plotted response-score convention, Energy Correction ($\epsilon_{back}$, $+2.52$) and Linear Drag Scale ($\alpha_r$, $-0.73$) produced the two strongest diagnostic responses, while Eddy Diffusivity ($K_{eddy}$) and the inactive Smagorinsky coefficient ($C_s$) produced negligible responses and were fixed in the hybrid search. For this case, the final GFO-MultiNN solution shows large adjustments in Enstrophy Correction ($\epsilon_{ens}$, $+127\%$), Drag Scale ($\alpha_r$, $+122\%$), Viscosity Scale ($\alpha_\nu$, $+98\%$), and the magnitude of Energy Correction ($\epsilon_{back}$, reduced by $86\%$). Across the full ensemble, the exact magnitudes vary, but the same small active subset and qualitative response structure recur. These associations suggest that energy dissipation and enstrophy regulation are important contributors to the depth-weighted diagnostic error in this model; they do not by themselves establish unique causal attribution.

This phase gives the hybrid workflow an interpretable screening layer that is not available from the surrogate predictions alone. GFO identifies which normalized-coordinate perturbations are most strongly associated with local loss changes and bounds the subsequent non-linear search accordingly. For the representative case, the responses implicate the balance among backscatter, drag, viscosity, and enstrophy correction as useful tuning directions. Because the probes are local and one-at-a-time, these findings are sensitivity-based physical diagnostics rather than proof that any one term is a unique causal source of error.

Notably, GFO-MultiGP (Gaussian Process) and GFO-MultiNN (Neural Network) achieved comparable final improvements of 64.6\% and 65.2\%, respectively, with similar convergence efficiency. Within the tested hybrid configuration, performance was therefore relatively insensitive to the choice between these two surrogate models. This comparison does not determine the separate contributions of screening, dimensionality reduction, initialization, or fidelity scheduling, but it suggests that either surrogate can operate effectively within the present pipeline.

Linear Scaling of Screening ($O(n)$): The computational cost of the sensitivity-based screening step scales linearly with the number of parameters, requiring only $n+1$ simulations to evaluate sensitivities for $n$ parameters. This is a significant advantage over uninformed sampling strategies, where adequate coverage of the parameter space grows combinatorially with dimension. We note that the GP surrogate's per-iteration training cost scales as $O(M^3)$ with the number of observed data points $M$, which is distinct from the parameter dimensionality $n$.

While the linear GFO limits screening cost to $n+1$ runs, for extremely large parameter sets (e.g., hundreds of parameters across holistic Earth System components), even $O(n)$ scaling may become a computational bottleneck. As an alternative first-pass screening procedure, one could employ the Simultaneous Perturbation Stochastic Approximation (SPSA) algorithm \citep{Spall1992}. SPSA perturbs all parameters simultaneously in random multivariate directions (typically $+1$ or $-1$ along normalized dimensions) and uses two objective evaluations to estimate a stochastic gradient for the full parameter vector. This yields an $O(1)$ number of model evaluations per gradient estimate with respect to parameter dimension, although multiple SPSA iterations are generally required and the resulting sensitivities can be noisy. In that setting, SPSA could serve as a coarse pre-screening or ranking tool before the more targeted hybrid refinement described here.

For GCMs with dozens of tunable parameters, we hypothesize that the hybrid strategy can remain tractable, provided that the effective dimensionality of the problem is low. Empirical evidence from weather prediction suggests that parameters governing short-term tendency errors often correlate with long-term climate statistics \citep{Rodwell2007, Ma2014}, which could enable screening via computationally inexpensive short integrations. By identifying the dominant parameters (e.g., $k \approx 5$--$10$) through such fast sensitivity probes, the search space could be reduced to a tractable manifold for the non-linear optimizer. Testing this hypothesis in a full GCM setting is a key direction for future work.

The approach assumes that high-impact parameters can be identified via linear perturbations. If an effect is purely non-linear, GFO may miss it. Moreover, because the unscreened baselines include the inactive $C_s$ dimension, part of the observed screening advantage may arise from removing this known null direction. A five-dimensional unscreened baseline with $C_s$ removed a priori is therefore needed to quantify the advantage when every supplied parameter affects the model. Together with component-level fidelity and initialization ablations, this is a priority for future work.

In addition to the Bayesian methods, we investigated other data-driven optimization strategies. We implemented Reinforcement Learning (RL) using Proximal Policy Optimization (PPO) with a Neural Network surrogate. However, this approach was found to be highly sample-inefficient compared to the hybrid strategy. We also attempted ``Reverse Modeling,'' training a model to map target error variables directly back to parameters. This inverse formulation struggled with the ill-posed nature of the mapping, resulting in slow convergence and high uncertainty in the parameter estimates.

The optimized parameters exhibit a finite spread across the 35 ensemble members. As described in Section~2.5, each member differs in the random phases and amplitude of its native-grid initial PV fields, so the initial state and adjustment pathway are not identical across the ensemble. The optimal parameter magnitudes are therefore not expected to be identical. Importantly, the spread is much larger in magnitude than in role: the same small subset of active parameters consistently emerges from GFO screening even though the best parameter values differ between members. The optimizer is thus not finding one universal parameter vector; it identifies common controlling directions and then tunes each ensemble realization. The detailed diagnostics in Section~3.3 correspond to one randomly selected member, while the basin comparisons in Section~3.4 indicate that at least part of the spread is systematically related to state-dependent sensitivity rather than solely to optimizer variability.

\section{Summary}

This Quasi-Geostrophic turbulence proof-of-concept shows that the tested pipeline combining linear physics-based screening, reduced-dimensional search, and non-linear multi-fidelity learning achieved the strongest end-to-end sample efficiency among the evaluated strategies. The present design does not isolate the contribution of any one component.

Specifically, among the methods tested here, the hybrid methods achieved the strongest overall end-to-end performance: GFO-MultiGP and GFO-MultiNN reached mean best improvements of 64.6\% and 65.2\%, respectively, compared with 55.9\% for MultiNN, 61.1\% for the standalone GP and 42.0\% for the standalone NN. At the same time, the hybrid pipeline reached practical saturation after only 2,520--3,060 simulation-days, compared with 3,600 for MultiNN and 6,660--7,740 for the standalone ML methods. This corresponds to 40.7--49.9 simulation-days per 1\% improvement for the full hybrid methods, versus 67.8 for MultiNN and 133.3--167.0 for standalone GP/NN, while validated standalone GFO saturates at 8.2\% improvement after 1,440 simulation-days (175.6 simulation-days per 1\% improvement). These comparisons establish an end-to-end advantage for the tested hybrid pipelines. Because the experiments change multiple components simultaneously, they do not determine the individual or interactive contributions of screening, dimensionality reduction, initialization, and multi-fidelity refinement.

Equally important, the multi-fidelity strategy remains useful, given the point-wise low-fidelity to high-fidelity correlation is modest. This is expected from nonlinear dynamics resolved in the QG and Earth system models. Rather than requiring exact agreement in absolute loss values, the method only needs the 30-day simulations to recover enough of the low-loss structure to guide the search toward promising regions. The positive rank consistency and 76\% promotion success rate indicate that the low-fidelity stage can screen candidates effectively within the present workflow, allowing the hybrid optimizer to approach a near-optimal basin in substantially fewer effective iterations before switching to the expensive 180-day refinement stage. At the same time, the modest correlation and the absence of naive promotion baselines mean that this evidence should be interpreted as supportive rather than definitive proof of fidelity transfer.

Finally, the identical-twin experiments identified multiple parameter configurations within 3\% of the best achieved loss, even when the initial guesses differed substantially. After excluding the inactive Smagorinsky coefficient from physical interpretation, the remaining spread is consistent with compensation among implemented dissipation and correction controls. Within this proof-of-concept, the result suggests that reproducing the target diagnostics may be more important than recovering one unique parameter vector. Whether comparable equifinality occurs in comprehensive Earth System Models remains a question for future work.

The comparable performance of GFO-MultiGP and GFO-MultiNN (64.6\% vs.\ 65.2\%) shows that both surrogate choices performed similarly within the tested hybrid configuration. Future work will extend this framework to full GCM parameter spaces and include component-level ablations to isolate the contributions of screening, dimension reduction, initialization, and multi-fidelity scheduling.

\bibliographystyle{plainnat}
\bibliography{references}

\appendix
\section{Appendix}

\subsection{Green's Function Optimization (Linear Stage)}
Our simplified Green's Function Optimization (GFO) stage, inspired by published Green's-function calibration methods \citep{menemenlis2005using, Strobach2022, Carroll2020}, serves as an initial linear probe for sensitivity-based parameter screening. Unlike the full weighted inverse formulations in those studies, the present implementation operates on a scalar loss. It assumes that the response of $\mathcal{L}(\theta)$ to a one-at-a-time parameter perturbation can provide a useful local ranking and that individual parameter effects are approximately independent near the default state.

For each of the $n=6$ parameters, we perform a one-at-a-time perturbation experiment. Given a default parameter state $\theta^{(0)}$, we perturb the $i$-th parameter by a fraction $\alpha$ (set to 50\%):
\begin{equation}
    \theta_i' = \theta_i^{(0)} (1 + \alpha)
\end{equation}
The 50\% perturbation is used for the primary GFO projection reported in the main results. As a perturbation-amplitude robustness check, we repeated the sensitivity screening with $\alpha=0.05$ and $\alpha=0.20$. Both tests recovered the same active-parameter subset and led to the same qualitative screening conclusion as the 50\% experiment. Thus, although the projected loss magnitude exhibits the non-linear saturation discussed in Section~3.1, the parameter-screening conclusion did not depend on the perturbation amplitude over the tested 5--50\% range.

We run a 180-day simulation with the perturbed set to obtain the loss $\mathcal{L}_i'$. Before computing the finite-difference sensitivity, each parameter is mapped to a common normalized coordinate:
\begin{equation}
z_i(\theta_i) =
\begin{cases}
\displaystyle \frac{\theta_i-\theta_{i,\min}}{\theta_{i,\max}-\theta_{i,\min}},
& \text{linear parameters},\\[1.1em]
\displaystyle \frac{\log_{10}(\theta_i)-\log_{10}(\theta_{i,\min}^{*})}
{\log_{10}(\theta_{i,\max})-\log_{10}(\theta_{i,\min}^{*})},
& \text{log-scaled parameters},
\end{cases}
\end{equation}
where $z_i$ is a dimensionless normalized coordinate. For a log-scaled parameter whose physical range includes zero, $\theta_{i,\min}^{*}$ is a strictly positive computational floor; for $\epsilon_{ens}$ we use $\theta_{i,\min}^{*}=10^{-10}\ \mathrm{s}^{-1}$. The sensitivity used for screening is then
\begin{equation}
    J_i^{(z)} \approx
    \frac{\mathcal{L}_i' - \mathcal{L}^{(0)}}
    {z_i(\theta_i')-z_i(\theta_i^{(0)}) + \varepsilon},
\end{equation}
with $\varepsilon=10^{-10}$ used only as a numerical safeguard. Thus, $J_i^{(z)}$ is the loss change per unit displacement across a common normalized coordinate, not a derivative with respect to a dimensional parameter. Ranking $|J_i^{(z)}|$ therefore does not compare quantities with incompatible physical units. The four parameters with the largest $|J_i^{(z)}|$ are selected as the active parameters for the subsequent non-linear optimization stage.

For the coarse GFO projection, each coordinate is updated independently using a damped and bounded gradient step,
\begin{align}
    \Delta z_i &= \operatorname{clip}\!\left(-\gamma J_i^{(z)},-\delta,\delta\right),\\
    z_i^{\mathrm{proj}} &= \operatorname{clip}\!\left(z_i^{(0)}+\Delta z_i,0,1\right),
\end{align}
where the implementation uses $\gamma=0.05$ and $\delta=0.2$. The projected coordinate is then mapped back to the physical parameter space using the inverse of the linear or logarithmic transformation above. This axis-aligned update deliberately omits cross-parameter interactions; its purpose is to provide a coarse initial projection and a scale-consistent sensitivity ranking before the subsequent non-linear refinement.

\subsection{Space-Filling Initialization}
To generate space-filling initial data, we use Latin Hypercube Sampling (LHS) \citep{McKay1979} for the Experiment 0 baseline and the 24-point initial designs in Experiments 2--3. LHS divides each parameter range into $M$ equal-probability intervals and draws one sample from each interval. The four-parameter hybrid warm start uses a mixed 16-evaluation design: one injected default reference, seven LHS points, and eight scrambled Sobol points. This mixed design provides both a known baseline anchor and complementary space-filling coverage.

We utilize the \texttt{scipy.stats.qmc.LatinHypercube} and \texttt{scipy.stats.qmc.Sobol} samplers. Sampling occurs in the unit hypercube $[0, 1]^D$:
1.   Generation:  Generate $M$ samples in the unit space $U \in [0, 1]^{M \times D}$.
2.   Warping:  Map the unit samples to physical parameter space using inverse Cumulative Distribution Functions (CDFs).
    \begin{itemize}
        \item For linear parameters (e.g., Viscosity, Drag): $P = P_{min} + U \cdot (P_{max} - P_{min})$.
        \item For logarithmic parameters (e.g., Eddy Diffusivity, Enstrophy Correction):
        \[
            P = 10^{\log_{10}(P_{min}) + U \cdot (\log_{10}(P_{max}) - \log_{10}(P_{min}))}
        \]
    \end{itemize}
This ensures that logarithmic parameters are explored across orders of magnitude. For the hybrid initialization, the seven LHS and eight scrambled Sobol points are mapped using the same transformations and followed by the small seeded design perturbation described in Appendix~A.3. Together with the default reference, they form the 16-evaluation warm start rather than 16 LHS points. For robustness, the execution is managed via a queue system that automatically retries simulations with new random samples if a run becomes numerically unstable (`NaN` loss).

\subsection{Neural Network Surrogate Model}
In Experiment 6 (GFO-MultiNN), we employ a Neural Network (NN) approximated via Monte Carlo (MC) Dropout \citep{Gal2016} to model the non-linear loss surface and its epistemic uncertainty. Because the GFO screening identified four active parameters ($\alpha_{\nu}$, $\alpha_{r}$, $\epsilon_{back}$, $\epsilon_{ens}$), the hybrid NN operates on this reduced input space; the remaining two parameters ($K_{eddy}$, $C_s$) are fixed at the values identified by that ensemble member's GFO screening throughout the hybrid optimization. The standalone Experiment 3 uses the same base architecture and MC-dropout uncertainty calculation but receives all six candidate inputs and is trained only on 180-day evaluations. Experiment 4 (MultiNN) is a distinct six-input multi-fidelity control implemented as a bootstrap ensemble of five \texttt{sklearn} \texttt{MLPRegressor} models with hidden layers $(64,32)$, ReLU activations, Adam optimization, an initial learning rate of $10^{-2}$ with adaptive scheduling, a 500-epoch maximum, and early stopping with a 0.1 validation fraction. Its predictive mean and ensemble spread are used by the same multi-fidelity acquisition workflow. Thus Experiment 4 tests the full unscreened MultiNN pipeline and is not presented as an architecture-matched component ablation of Experiment 6.

Architecture:
The model is a fully connected Multi-Layer Perceptron (MLP) implemented in TensorFlow/Keras:
\begin{enumerate}
    \item  Input Layer:  4 neurons (the active normalized parameters; 6 neurons for standalone Experiment 3).
    \item  Hidden Layers:  3 dense layers with 64 neurons each, using Rectified Linear Unit (ReLU) activation.
    \item  Dropout:  A dropout rate of $p=0.1$ is applied after each hidden layer. For uncertainty-aware acquisition, dropout is explicitly enabled at inference.
    \item  Output Layer:  1 linear neuron predicting the loss metric.
\end{enumerate}

The input parameters are supplied in the normalized coordinates described in Appendix~A.1: linearly bounded parameters are mapped to $[0,1]$, while log-scaled parameters are transformed in base-10 logarithmic space and then mapped to $[0,1]$. At every surrogate update, the loss targets in the complete set of finite observations available at that iteration are standardized as $y^{*}=(y-\mu_y)/(\sigma_y+10^{-8})$; predictions are transformed back to physical loss units before acquisition. The network is compiled with the Adam optimizer, a learning rate of $10^{-2}$, and mean-squared-error training loss. Each update permits at most 200 epochs with a batch size of $\min(32,n_{train})$. Early stopping monitors the training loss with patience 20 and restores the best weights. No validation split is used. The model object is created once, and each optimization iteration fits it again on the full accumulated finite dataset while retaining the weights from the previous fit; this is therefore sequential warm-start retraining rather than initialization of a new network at every iteration. NumPy and TensorFlow are both initialized with algorithmic seed 42 in the reported runs. The LHS and scrambled Sobol initialization use seeds 42 and 1042, respectively, followed by a small Gaussian design perturbation generated from seed 2042.

Uncertainty Estimation (MC Dropout):
To capture epistemic uncertainty, we perform $T=50$ forward passes for each prediction input $x^*$, identifying the mean $\mu(x^*)$ and standard deviation $\sigma(x^*)$ of the stochastic outputs:
\begin{equation}
        \mu(x^*) \approx \frac{1}{T} \sum_{t=1}^T f(x^*; W_t), \quad \sigma^2(x^*) \approx \frac{1}{T} \sum_{t=1}^T (f(x^*; W_t) - \mu(x^*))^2
\end{equation}
where $W_t$ represents the randomized network weights due to dropped connections. The implementation uses $T=50$ and takes the empirical standard deviation of these predictions; it does not add a separately estimated observation-noise variance. This uncertainty estimate allows the optimizer to employ Thompson Sampling and Upper Confidence Bound (UCB) acquisition strategies.

\subsection{Gaussian Process (GP) Ensemble}
For Experiments 2 (standalone GP) and 5 (GFO-MultiGP), we utilize an ensemble of Gaussian Processes \citep{Rasmussen2006} to model the objective function probability distribution $P(f|D)$. Experiment 2 operates in the full six-dimensional parameter space using only 180-day evaluations, whereas Experiment 5 operates in the four-dimensional GFO-selected subspace under the multi-fidelity schedule. A GP is defined by a mean function $m(x)$ and a covariance kernel $k(x, x')$.

To make minimal assumptions about the smoothness of the loss landscape, we construct an ensemble of $K=8$ distinct kernels using `sklearn.gaussian\_process`:
\begin{equation}
    k_{ens}(x, x') = \sum_{j=1}^{K} w_j \cdot k_j(x, x')
\end{equation}
The ensemble includes:
\begin{itemize}
    \item  Matérn 1.5 \& 2.5:  Handling rough to moderately smooth landscapes.
    \item  Radial Basis Function (RBF):  Handling smooth, infinitely differentiable features.
    \item  Varying Length Scales:  Ranging from 0.3 (local fit) to 2.0 (global trend).
\end{itemize}
The weights $w_j$ are determined dynamically based on the log-marginal likelihood of each kernel given the observed data. The final prediction uses a conservative uncertainty estimate by taking the maximum variance across the weighted ensemble to prevent over-confidence in unexplored regions.

\subsection{Optimization Heuristics \& Helper Functions}
Several heuristic strategies were implemented to enhance the robustness and efficiency of the shared hybrid optimization loop (Experiments 5 and 6). Unless otherwise noted, these heuristics are applied in the same way for both GFO-MultiGP and GFO-MultiNN; the primary difference between the two methods lies in how surrogate uncertainty is estimated.

1. Trust Region Optimization:
To prevent the optimizer from oscillating wildly in the high-dimensional space, we constrain the search to a dynamic Trust Region \citep{Conn2000} (hyper-rectangle) centered around the best-known solution $x_{best}$.
\begin{itemize}
    \item  Expansion:  If a new best solution is found for 3 consecutive iterations, the radius $R$ expands ($R \leftarrow \min(1.0, R \cdot 1.5)$).
    \item  Contraction:  If no improvement is found for 3 iterations, $R$ contracts ($R \leftarrow \max(0.05, R \cdot 0.5)$) to focus on local refinement.
\end{itemize}

2. Hybrid Acquisition Function:
We select candidates by maximizing a hybrid utility function combining Exploration (UCB) and Exploitation (Expected Improvement, EI) \citep{Jones1998, Srinivas2010}, augmented with Local Penalization (LP) \citep{Gonzalez2016} to prevent redundant sampling:
\begin{equation}
    a(x) = \left[ w_{EI} \cdot EI(x) + (1-w_{EI}) \cdot UCB(x) \right] \cdot \prod_{j} (1 - \text{Pen}(x, x_j))
\end{equation}
where $\text{Pen}(x, x_j)$ penalizes regions close to previously sampled points $x_j$, forcing the optimizer to explore elsewhere.

3. Thompson Sampling:
For randomized exploration--exploitation phases, we employ Thompson Sampling \citep{Thompson1933, chapelle2011empirical}. We draw a randomized function realization from the current surrogate predictive distribution and select the minimizer of this stochastic function. In Experiment 5, this realization is taken from the GP ensemble posterior; in Experiment 6, it is generated from the MC-Dropout NN predictive ensemble. This posterior-sampling strategy provides stochastic exploration without relying solely on a deterministic UCB score.

\subsection{QG Numerical Configuration and Initialization}
\label{app:qg_impl}
Both resolutions use a Fourier pseudospectral discretization on a doubly periodic $L_x\times L_y=2.0\times10^6\ \mathrm{m}\times1.0\times10^6\ \mathrm{m}$ domain. The HR grid is $512\times256$ with $\Delta t=600\ \mathrm{s}$, and the LR grid is $64\times32$ with $\Delta t=1800\ \mathrm{s}$. The shared physical configuration is
\begin{equation}
    \beta=1.8\times10^{-11}\ \mathrm{m}^{-1}\mathrm{s}^{-1},\quad
    f_0=10^{-4}\ \mathrm{s}^{-1},\quad g'=0.02\ \mathrm{m}\,\mathrm{s}^{-2},\quad
    H_1=750\ \mathrm{m},\quad H_2=3250\ \mathrm{m},
\end{equation}
with base lower-level drag $r_{ek}=8\times10^{-8}\ \mathrm{s}^{-1}$ and base spectral-filter coefficient $\nu=3\times10^{14}$. These values give the configured diagnostic deformation wavenumber $k_d=2.86\times10^{-5}\ \mathrm{m}^{-1}$ ($k_d^{-1}\simeq34.9\ \mathrm{km}$), although, as noted in Section~2.1, $k_d$ is not used by the archived inversion.

Spatial derivatives and the Jacobian are evaluated spectrally. Fourier coefficients outside the two-thirds mask are removed when constructing the Jacobian and after every time step. Time integration uses forward Euler for the first step, second-order Adams--Bashforth for the second, and third-order Adams--Bashforth thereafter. Following each explicit update, both PV fields are multiplied by
\begin{equation}
    F_{\nu}(\boldsymbol{k})=
    \exp\!\left[-\alpha_{\nu}\nu\Delta t\left(|\boldsymbol{k}|^2\right)^4\right].
\end{equation}
For LR runs with $K_{eddy}>0$, the additional multiplier is
\begin{equation}
    F_K(\boldsymbol{k})=
    \exp\!\left[-K_{eddy}\Delta t\left(|\boldsymbol{k}|^2\right)^2\right].
\end{equation}
Thus the implemented viscosity and eddy-diffusion controls are eighth-order and fourth-order spectral damping, respectively; no additional time filter is applied. Lower-level drag enters the resolved tendency as $\alpha_r r_{ek}\nabla^2\psi_2$. The enstrophy correction is $-\epsilon_{ens}q_i$, and stochastic correction draws are independent across grid points, levels, and time steps with tendency amplitude $10^{-8}\epsilon_{back}$.

Initial PV is generated independently on each numerical grid as a sum of 12 Gaussian vortices plus two scales of Gaussian noise. The four largest vortices have centers $(0.2,0.3)$, $(0.8,0.7)$, $(0.2,0.7)$, and $(0.8,0.3)$ in units of $(L_x,L_y)$; widths $(0.13,0.13,0.12,0.12)L_x$; upper-level amplitudes $(4,-4,3.5,-3.5)\times10^{-6}\ \mathrm{s}^{-1}$; and lower-level amplitudes $(2.5,-2.5,-2,2)\times10^{-6}\ \mathrm{s}^{-1}$. The remaining eight vortices are specified by the tuples $(x/L_x,y/L_y,\sigma/L_x,10^6A_1,10^6A_2)$:
\begin{equation}
\begin{split}
 &(0.35,0.50,0.09,3,1.5),\ (0.65,0.50,0.09,-3,-1.5),\\
 &(0.50,0.25,0.08,2.5,0),\ (0.50,0.75,0.08,-2.5,0),\\
 &(0.40,0.35,0.06,2,1),\ (0.60,0.65,0.06,-2,-1),\\
 &(0.30,0.60,0.05,1.5,0.8),\ (0.70,0.40,0.05,-1.5,-0.8).
\end{split}
\end{equation}
Ensemble-specific seeds displace the four large-vortex centers by independent uniform offsets in $[-0.1L_x,0.1L_x]$ and $[-0.1L_y,0.1L_y]$. The added noise uses native-grid standard-normal fields at scales 1 and 2, with amplitudes $8\times10^{-7}/(s+1)$ for scale $s$; the scale-2 field is bilinearly interpolated to the model grid. HR and LR initial fields are therefore generated from the same member seed and analytic prescription, but are not obtained by exact spectral truncation of one common realization. The default LR controls are $(\alpha_\nu,\alpha_r,K_{eddy},C_s,\epsilon_{back},\epsilon_{ens})=(0.5,0.5,10^3,0.015,-0.002,3\times10^{-9})$.

The integrator checks both PV fields for NaN or infinite values after every step and terminates an unstable integration immediately. The optimization wrappers catch simulation exceptions and non-finite objective values, record the evaluation as invalid, exclude it from surrogate fitting, and propose a replacement candidate. The QG fields are archived every 12~h in the surrogate experiments (24~h for the GFO screening runs); the time means in Section~2.2 are discrete averages over those saved fields.

\clearpage
\section*{Supplementary Material}
\setcounter{figure}{0}
\renewcommand{\thefigure}{S\arabic{figure}}

\begin{figure}[H]
    \centering
    \includegraphics[width=0.9\textwidth]{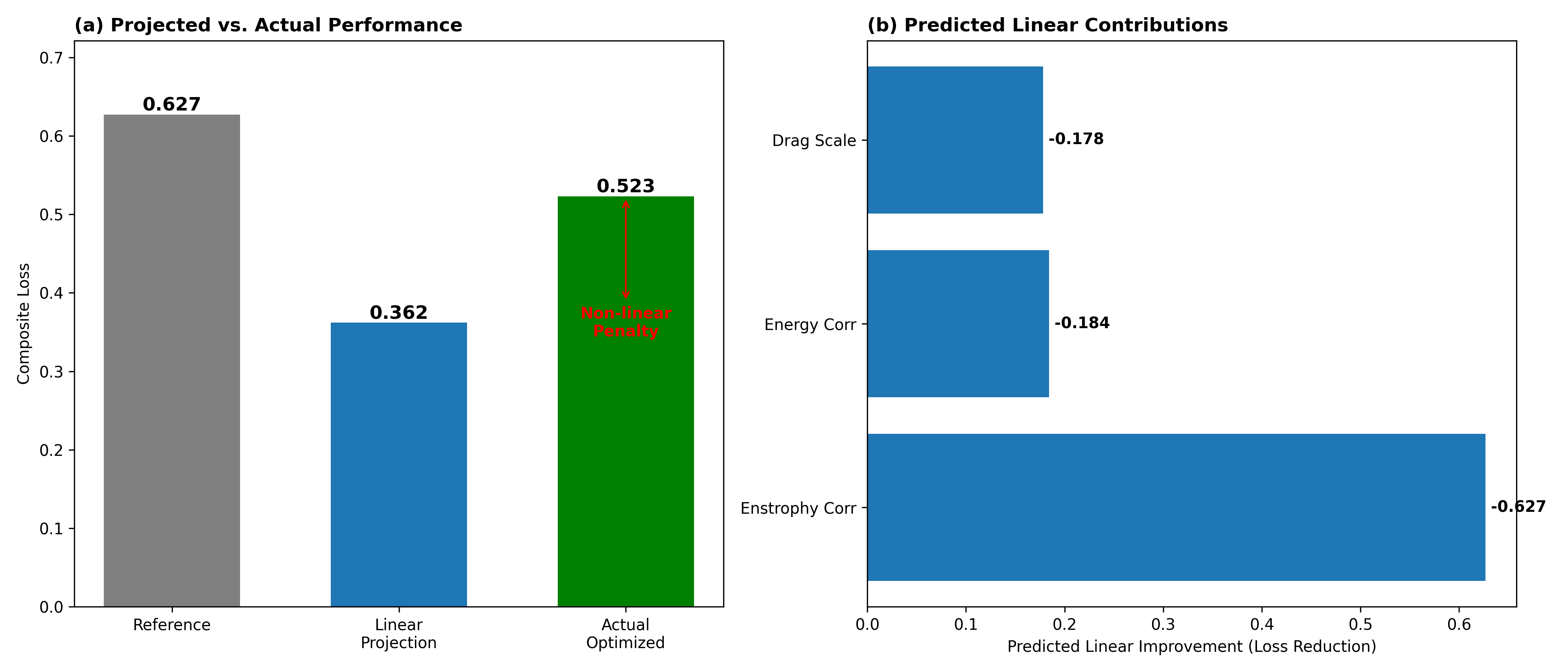}
    \caption{Projected vs. Actual Improvement. (a) The linear model projected a cost drop to 0.429 (Projected), whereas the actual non-linear system saturated at 0.461. (b) The discrepancy illustrates the ``Non-linear Penalty'' or diminishing returns inherent in turbulent optimization, highlighting that while linear sensitivity identifies the correct descent direction, it cannot accurately predict large-step magnitudes.}
    \label{fig:gf_projection_supp}
\end{figure}

\begin{figure}[H]
    \centering
    \includegraphics[width=0.9\textwidth]{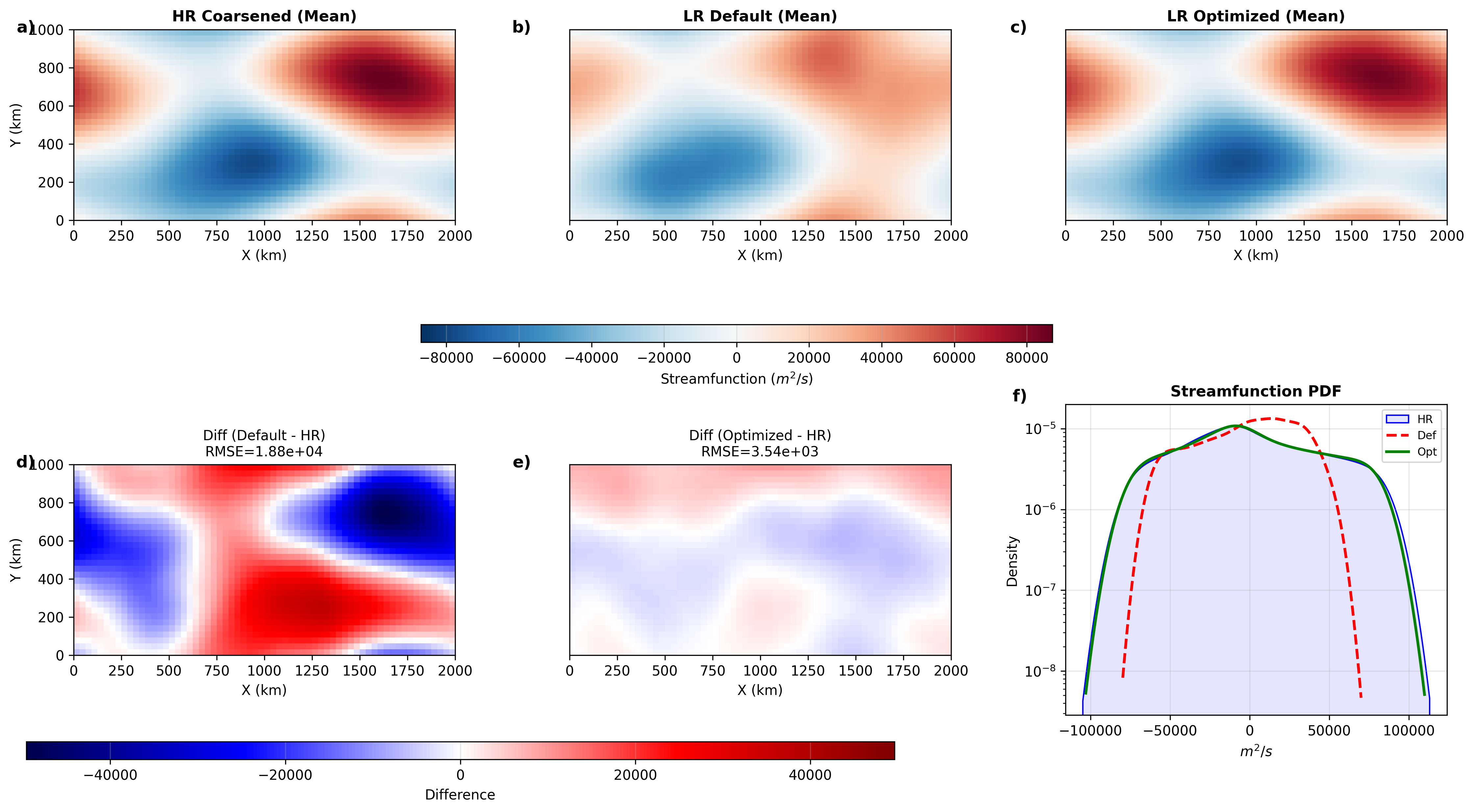}
    \caption{Streamfunction ($\psi$) Field Comparison. Top: High-Resolution Reference. Middle: Default Low-Resolution. Bottom: Optimized Low-Resolution. The optimized model effectively corrects the zonal jet structure and reduces large-scale biases.}
    \label{fig:field_psi}
\end{figure}

\end{document}